\documentclass[10pt,letterpaper]{article}

\usepackage[latin1]{inputenc}
\usepackage{amsmath}
\usepackage{amsfonts}
\usepackage{amssymb}
\usepackage{graphicx}
\usepackage{fullpage}
\usepackage{float}
\usepackage{multicol,caption}
\usepackage[parfill]{parskip} 
\usepackage{color}
\usepackage[colorlinks=true, pdfstartview=FitV, linkcolor=blue,citecolor=blue, urlcolor=blue]{hyperref}
\usepackage{osajnl2}
\usepackage{epstopdf}
\usepackage{booktabs}
\usepackage{ctable}
\usepackage{array}
\usepackage{subfig}
\usepackage{cite}

\begin{document}

\title{Intermodal Energy Transfer in a Tapered Optical Fiber: Optimizing Transmission.}

\author{S. Ravets,$^{1,2}$ J. E. Hoffman,$^1$ P. R. Kordell,$^1$ J. D. Wong-Campos,$^1$ S. L. Rolston,$^1$ and L. A. Orozco$^{1,*}$}

\address{
$^1$Joint Quantum Institute, Department of Physics, University of Maryland, and National Institute of Standards and Technology, College Park, MD 20742, U. S. A.\\
$^2$ Laboratoire Charles Fabry, Institut d'Optique, CNRS, Univ Paris Sud, 2 avenue Augustin Fresnel, 91127 Palaiseau cedex, France\\
$^*$Corresponding author: lorozco@umd.edu
}
\date{today}

\begin{abstract*}
We present an experimental and theoretical study of the energy transfer between modes during the tapering process of an optical nanofiber through spectrogram analysis. The results allow optimization of the tapering process, and we measure transmission in excess of 99.95\% for the fundamental mode. We quantify the adiabaticity condition through calculations and place an upper bound on the amount of energy transferred to other modes at each step of the tapering, giving practical limits to the tapering angle. 
\end{abstract*}

\section{INTRODUCTION}
The study of mode coupling in an optical waveguide \cite{Yariv1973} is fundamentally important for good control of connectorization and transmission. This is especially true for tapered optical fibers with sub-wavelength waists, where light propagates in a mode that exhibits a large evanescent component propagating outside the waveguide. Nanofibers are ideal for probing nonlinear physics, atomic physics, and other sensing applications \cite{Kien2004, LeonSaval2004, Vetsch2010, Bures1999}. As the light propagates through the taper, it successively encounters regimes where the fiber is single mode, multimode and then single mode again. Careful design of adiabatic tapers leads to ultra-low loss fibers \cite{Tong2003}. Adiabatic criteria give an upper limit on how steep a taper can be, but are too vague for optimization of transmission. Here we are interested in giving quantitative bounds and constraints on the taper geometry.

Using a spectrogram analysis of the transmission signal through the fiber \cite{Orucevic2007}, we are able to identify the modes excited during the tapering process and extract the amount of energy transferred to each of these modes. Using this analysis, we show the importance of the geometry control and the fiber cleanliness to reach transmissions as high as 99.95\% in commercial fibers at 780 nm. Our nanofibers can handle more than 400 mW of optical power in ultra-high vacuum. After reaching the cutoff radius, the excited modes couple to radiative modes \cite{Snyder1983} and diffract outside of the fiber. 
 
Our analysis provides a path to fully model the electromagnetic field evolution in a nanofiber. This is crucial for a complete modeling of the coupling between light and matter \cite{Balykin2004, Spillane2003}. In the example of atoms trapped on the evanescent field around a nanofiber waist, it is necessary to know the coupling coefficients between the modes of the field and the atoms. This work details the modal evolution in the fiber, opening perspectives for the design of even more adiabatic fibers, making them usable in extreme conditions \cite{Hoffman2011}.

This paper presents our protocols, diagnostics, and characterization tools for fabricating nanofibers. It is structured as follows: we first overview our experimental goals and conditions in Sec.~\ref{sec:motivation}. Section~\ref{sec:modal} presents the modal evolution in tapered fibers. We then study in Sec.~\ref{sec:adiabaticity} adiabaticity in tapered fibers. Section~\ref{sec:transanalysis} analyzes in more detail the transmission signal. We introduce the spectrogram to analyze the transmission \cite{Orucevic2007} in Sec.~\ref{sec:spectrogram}. By modeling and diagnosing the fiber pull, we identify in Sec.~\ref{sec:applications} crucial elements to improve the transmission. Section~\ref{sec:thelosses} looks into the other losses present in the fiber. Section~\ref{sec:conclusion} is the conclusion of the paper.

\section{MOTIVATION AND CONSTRUCTION OVERVIEW}
\label{sec:motivation}

Controlling neutral atoms with dipole traps is a successful and promising avenue for the implementation of a growing number of scientific and technical applications \cite{Grimm2000}. The off-resonant interaction between light and atoms in the presence of an intensity gradient produces a dipole force that can generate traps: detuning below atomic resonance attracts atoms to go to the most intense region creating an optical tweezer \cite{Ashkin1978,Chu1986b} and above resonance detuning keeps the atom in the intensity minima, requiring more complicated geometries \cite{Cook1982,Yang1986,Davidson1995,Kulin2001}. One drawback of optical tweezers obtained by tightly focusing a laser beam comes from diffraction, which limits the trapping volume extension in the axial direction. One solution to this limitation is the use of optical nanofibers \cite{Balykin2004,Vetsch2010,Goban2012}. These devices offer enough light confinement and guidance to trap atoms over a few centimeters in the axial direction and present the advantage of being integrable to other devices \cite{Spillane2003,Jiang2006,Nayak2011,Wuttke2012}. We are interested in introducing this device into a 12~mK cryogenic environment to probe interactions between a trapped neutral atom and a superconducting circuit \cite{Hoffman2011}. 

Following the work of Warken et al. in \cite{Warken2007b}, we produce our fibers using a heat-and-pull technique, summarized below (see Ref.~\cite{Hoffman2013} for details on the algorithm and the hardware). An oxyhydrogen flame at stoichiometric combination brings a 0.75~mm long fiber portion to a temperature that exceeds its softening point. Two high-precision computer-controlled motors pull on the fiber ends at a typical velocity of 0.1~mm/s. We use an algorithm that relies on conservation of volume, which calculates the trajectories of the motors to produce a fiber of chosen geometry. The code is available at the DRUM Digital Repository of the University of Maryland \cite{DRUM}.

We pull a SM800 fiber from Fibercore that has a numerical aperture of 0.12 and a cutoff wavelength of 794~nm. Using the Sellmeier coefficients provided by Fibercore, we determine the core ($n_{core} = 1.45861$) and the cladding ($n_{clad} = 1.45367$) indices of refraction. The pull is divided into approximately 100 steps, such that the taper is composed of a series of sections small enough to be considered linear. Our tapers are generally composed of a section with a constant few mrad taper angle that reduces the fiber to a radius of 6~$\mu$m, and then connects to an exponential section that gently reaches submicrometer radii (on the order of 250~nm). The central waist is uniform and its length can be between 5~mm and 10~cm. A pull generally lasts for a few hundreds of seconds.

\section{MODAL EVOLUTION}
\label{sec:modal}

\subsection{Modes in a cylindrical waveguide}
\label{sec:modes}

The description of modes in a cylindrical waveguide using Maxwell equations can be found in several references {\it e. g.}\cite{Yariv1990,Snyder1983}. The modal fields vary as $\exp{\left[i(\beta_{lm}z-\omega t)\right]}$ where $\beta_{lm}$ is the propagation constant of the mode of symmetry and order $(l,m)$. The propagation of light inside a two-layered step index fiber depends on the $V$-parameter of the fiber, 

\begin{equation}
V=\frac{2\pi}{\lambda}a\sqrt{n_1^2-n_2^2},
\label{V}
\end{equation}

where $a$ is the core radius, $n_1$ is the core index of refraction, $n_2$ is the surrounding medium index of refraction, and $\lambda$ is the free space wavelength. The relation between $\beta_{lm}$ and the $V$-parameter is called the dispersion relation of mode $(l,m)$. In our tapers, we can approximate the fiber as a two-layer step index cylindrical waveguide in two regions: at the beginning of the taper, the light is confined to the core and guided through the core-to-cladding interface. We assume that the core and the cladding radii decrease at the same rate along the taper, which implies that there is no diffusion of the core into the cladding during the tapering process. In the waist, what was initially the core in the center of the fiber is now negligible ($a_{core} \approx 10~\rm{nm} \ll \lambda$). The light is then guided through the cladding-to-air interface.

\begin{figure}[H]
\centering
\includegraphics[width=0.7\linewidth]{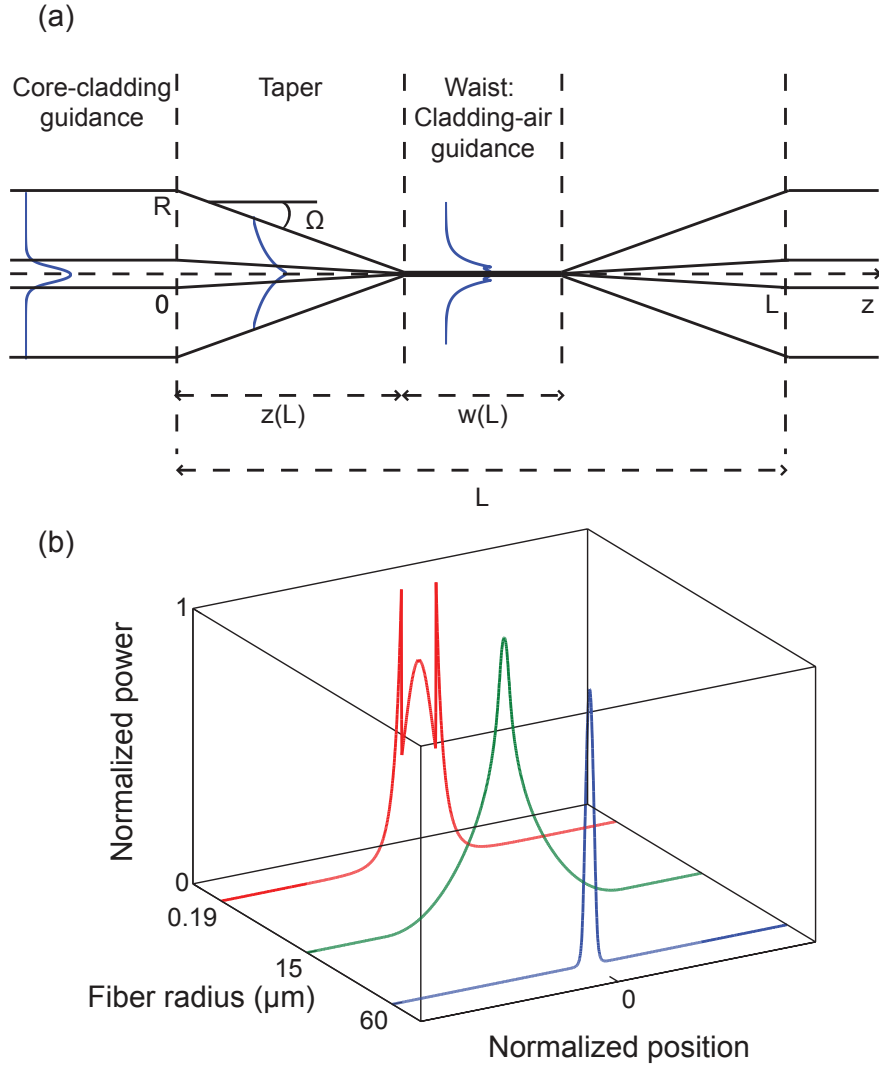}
\caption{(a) Schematic of the stretched fiber. At a given time, the fiber is composed of two tapers and an uniform waist of radius $r$ and length $w$. The total stretch is equal to $L$. (b) Calculated intensity profile of the mode for a radius of the fiber equal to 60~$\mu$m, 15~$\mu$m and 190~nm. Note that the position axes are not quantitative, and have been scaled to make the plots visible. The profiles are normalized to their maximum power.}
\label{fig:fiber}
\end{figure}

\subsection{Three-layer fiber}

\begin{figure}[H]
\label{dispersion}
\centering
\includegraphics[width=0.8\linewidth]{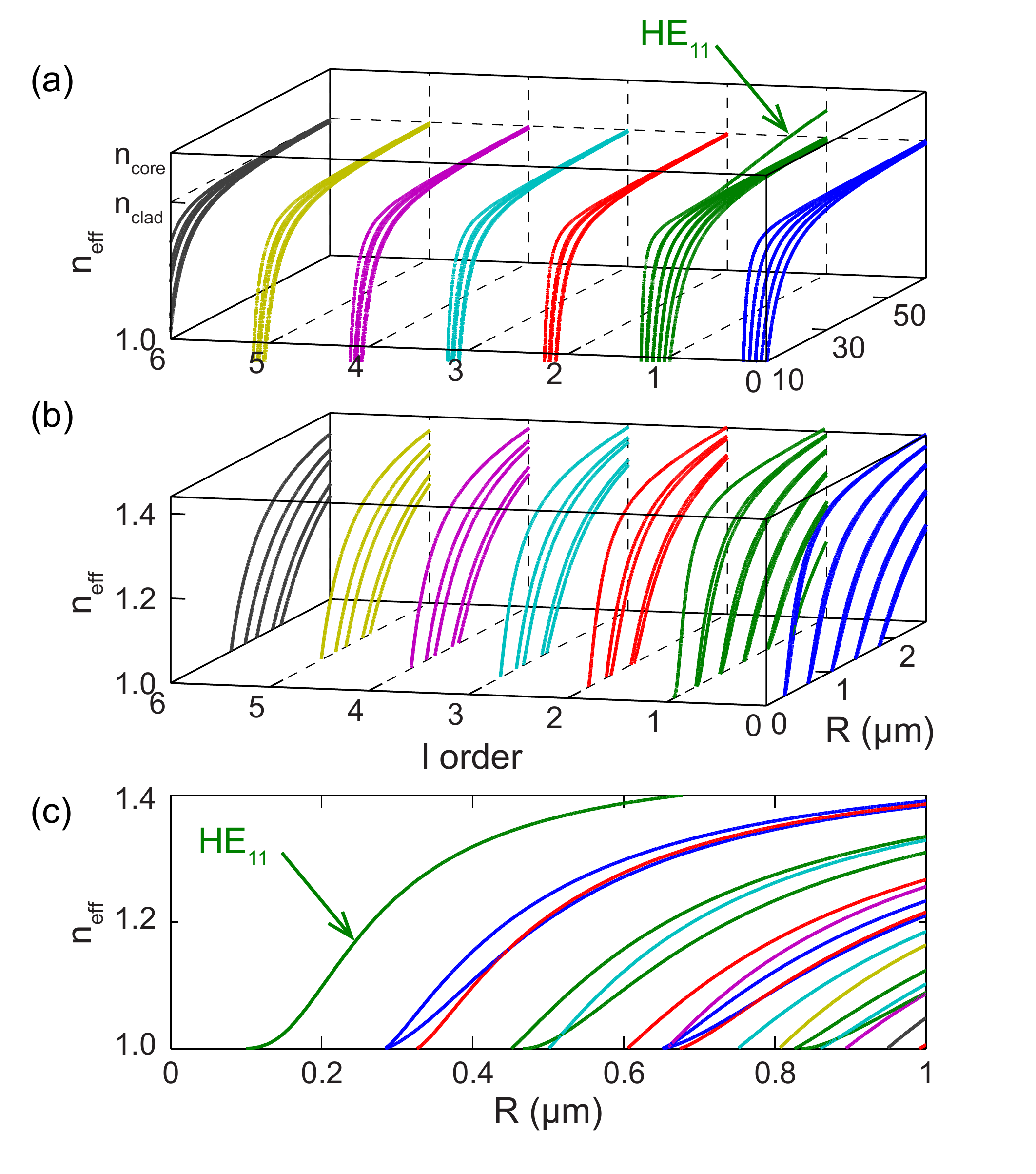}
\caption{(color online) Dispersion relations for various modes $TE_{0m}$, $TM_{0m}$, $HE_{lm}$ and $EH_{lm}$ ($l=1~{\rm{to}}~5$) as a function of radius calculated for a three-layer model using FIMMPROP. Here, $n_{air}=n_{vacuum}=1$. We show the first few modes of each family. (a)-(b) Three dimensional representation of the dispersions for different mode families. (c) Projection for the smaller values of R.}
\label{fig:dispersion}
\end{figure}

Since we continuously decrease the fiber radius during the pull, the fundamental mode leaks from the core to the cladding. In that region, the presence of the core, the cladding, and the air influence the mode (see Fig.~\ref{fig:fiber}). A proper treatment has to take into account all of those interfaces. We model our fibers by a three-layered structure, and we calculate the dispersion relations for a series of modes using the fully vectorial finite difference mode solver from commercial software FIMMWAVE \cite{FIMMPROP}. Figure~\ref{fig:dispersion} shows a plot of $n_{\rm{eff}} = \beta/k_0$ as a function of the radius of the SM800 fiber described in Sec.~\ref{sec:motivation}.

We are interested in modes that are initially launched into the core, thus guided by the core-to-cladding interface. Core modes have most of their energy contained in the core, and their effective indices of refraction satisfy $n_{clad}<n_{eff}<n_{core}$. Figure~\ref{fig:dispersion} shows that the $HE_{11}$ mode effective index is initially greater than $n_{clad}=1.45367$ (green curve indicated by an arrow). Some higher-order modes from LP11 family may be accepted in the core, close to their cutoff condition (the fiber cutoff wavelength is 792~nm~$>$~780.24~nm, so strictly speaking, we are not working in the fiber single mode regime). Experimentally, we filter higher order modes that have been launched or excited with a 1.27~cm diameter mandrel, effectively placing us into the single mode regime.

When the fiber radius decreases, $n_{eff}^{HE11}$ approaches $n_{clad}$. Since we model the fiber by a three-slab cylindrical waveguide, the cladding area is finite: the core becomes too small to support the fundamental mode around the point where $n_{eff}^{HE11}$ reaches $n_{clad}$ ($R = 19.43$~$\mu$m in Fig.~\ref{fig:dispersion}). The mode progressively leaks into the cladding to be guided by the cladding-to-air interface. The characteristic length-scale of the waveguide is $R \gg \lambda$, and many modes can be guided by the cladding to air interface ($n_{air}<n_{eff}<n_{clad}$), together with the fundamental mode. As long as $R \gg \lambda$, the air has little influence on the effective index of many of the accepted modes ($n_{eff} \approx n_{clad}$ for all the modes shown in Fig.~\ref{fig:dispersion}). The indices are so close to each other that the modes interact and exchange energy easily. For that reason, this is the critical region of the taper, where the adiabaticity condition is the most stringent. By symmetry, for a fully cylindrical fiber intermodal energy transfer will only happen between modes of the same family (one color in Fig.~\ref{fig:dispersion}). Energy transfers between modes from different families are a consequence of the presence of asymmetries.

Further decreasing $R$, we observe that the modes effective indices approach $n_{air}=1$. The dispersion curves separate, and adiabaticity can again be easily achieved. When the index of refraction of a mode reaches $n_{air}$, the mode is not guided by the fiber anymore and radiates into the air. This radius, specific to each mode, is called its cutoff. The highly-excited modes leave the fiber first, and the number of modes allowed in the waveguide decreases progressively (see Fig.~\ref{fig:dispersion}(c)). Under 0.3~$\mu$m, the only mode that can propagate is the $HE_{11}$ mode, whose index asymptotically approaches 1. The fiber is once again single-mode.

\section{ADIABATICITY IN FIBERS}
\label{sec:adiabaticity}

Achieving high transmission in nanofibers requires precise control of the taper geometry, where the mode adiabatically escapes from the core to the cladding before coupling back to the core \cite{Birks1992,Snyder1983}. High transmission through tapered nanofibers is indicative of their quality \cite{Tong2003,Fujiwara2011}. 

\subsection{Adiabaticity criterion}
\label{sec:Snydercriteria}

The mode conversion in a taper is strongly related to the taper geometry. If a taper is too short (taper angle too steep), the mode evolution is non-adiabatic, and we observe losses. Inversely, as the taper is lengthened, the mode conversion is more adiabatic. In the limit of a very shallow angle we intuitively understand that the transmission can reach 100\%, since all the energy remains in the fundamental mode throughout the evolution. Following this idea, an adiabaticity criterion has been derived \cite{Snyder1983} relating the characteristic taper length $z_t$, to the characteristic beating length between two modes $z_b$.

$z_t$ is the length associated with the tapering angle $\Omega$ at radius $R$, defined by:

\begin{equation}
\label{eq:taperlength}
z_t =\frac{R}{\tan(\Omega)},
\end{equation}

$z_b$ is the beat length between two modes (the spatial frequency of the beating):

 \begin{equation}
 \label{eq:beatlength}
z_b =\frac{2 \pi}{\beta_1 - \beta_2} = \frac{\lambda}{n_{eff,1}-n_{eff,2}}.
\end{equation}

where $\beta_1$ is the fundamental mode propagation constant at radius $R$ and $\beta_2$ is the propagation constant at radius $R$ of the first excited mode with the same symmetry as the fundamental mode ($EH_{11}$). Equation~(\ref{eq:beatlength}) relates the beat length to the inverse of the distance between two curves in Fig.~\ref{fig:dispersion}. Mode conversion in a taper is adiabatic when the fiber is long enough: $z_t \gg z_b$ \cite{Snyder1983}. If the two modes are close, $z_b$ is large, making the adiabaticity condition more difficult to satisfy. The choice of $EH_{11}$ gives the most stringent condition on the fiber length, as it produces the shortest beat length between the fundamental mode and any mode with symmetry $l=1$. Nevertheless, this condition remains too vague when one wants to optimize the taper geometry for a given transmission.

\begin{figure}[H]
\centering
\includegraphics[width=0.8\linewidth]{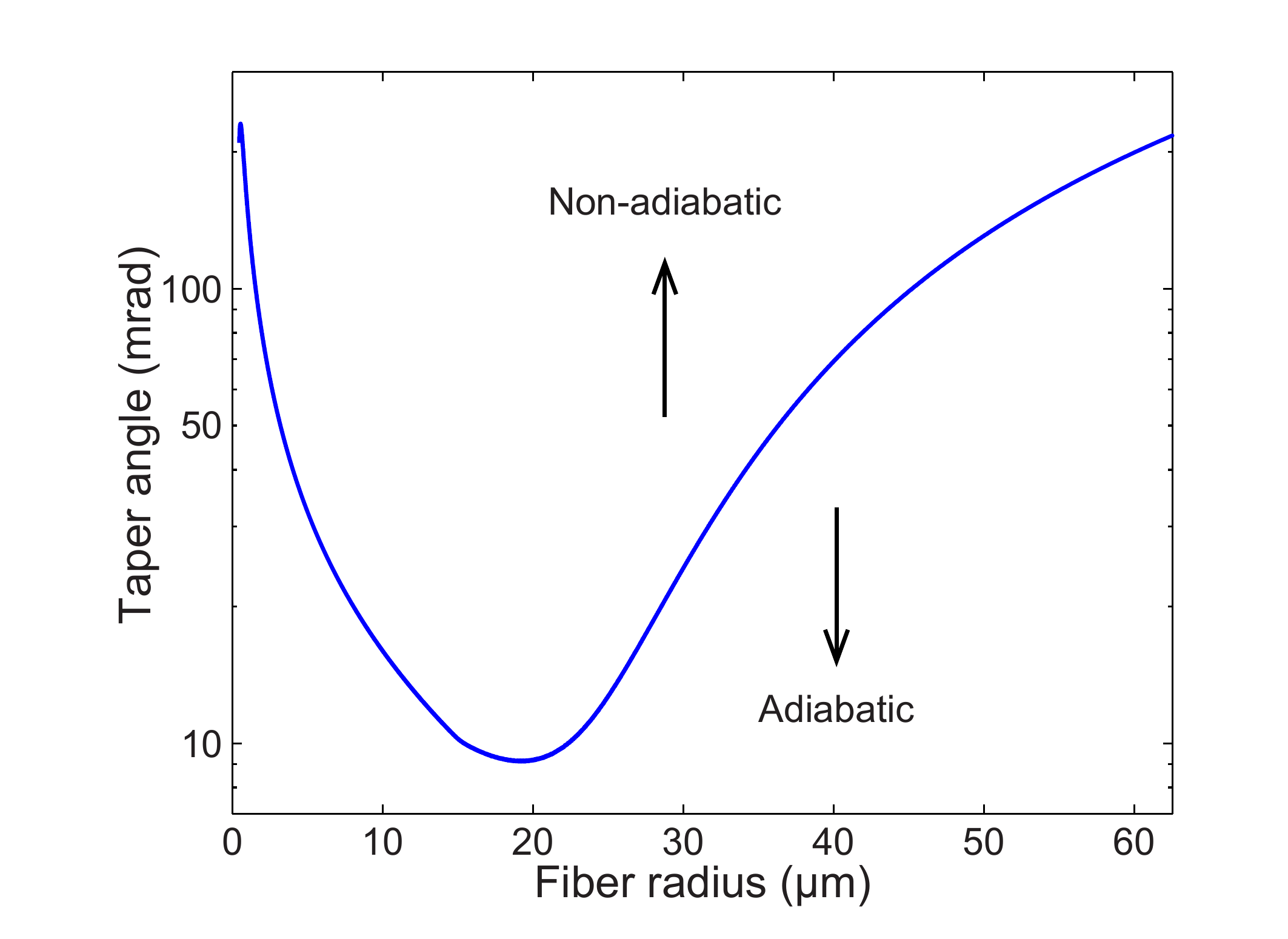}
\caption{(color on line) Upper boundary for the taper angle $\Omega$ as a function of the radius of the fiber set by $z_b=z_t$. Note the logarithmic scale for the vertical axis. The core to cladding diameter ratio for this fiber is 2.535/62.55 is fixed for the entirety of the pull. $n_{core} = 1.45861$ and $n_{clad} = 1.45367$.}
\label{fig:adiabaticity}
\end{figure}

Using the dispersion relations from FIMMPROP, we can solve the equation $z_t=z_b$, when the beat length equals the taper length. The blue curve in Fig.~\ref{fig:adiabaticity} separates the plane into two regions: in order to be adiabatic, taper angles need to be much smaller than the ones indicated on the curve. Above the curve, the angles correspond to non-adiabatic tapers. 

Figure~\ref{fig:adiabaticity} gives an \textit{upper} limit on the taper angle at a specific radius using the condition $z_b=z_t$ from Eqs.~\ref{eq:taperlength},~\ref{eq:beatlength}. It does not provide any quantitative information on the intermodal energy transfers for a given taper: calculations in Sec.~\ref{sec:fullyadiab} show that the angles in Fig.~\ref{fig:adiabaticity} lead to large anergy transfers. We are interested in producing fibers with high transmissions, greater than 99.90\%, and we need to find the optimal geometry necessary to reach a specific transmission. 

\subsection{Transmission of a tapered fiber section}

We perform numerical simulations with FIMMPROP to explore the parameter space and find the optimal adiabatic profile for a given transmission. The fiber tapers from 62.55~$\mu$m radius down to 250~nm radius. Using the indices of refraction for our SM800 fiber (see Sec.~\ref{sec:motivation}). We divide the taper into a discrete series of linear sections (32 sections in the present work). At the end of each section we project the output field into the first family of modes (here we use the 15 first modes of family 1) to obtain specific amplitude and phase information in terms of the excited modes. The $S$-matrix, relating input and and output, contains all the mode phases and amplitudes necessary to relate the input and output fields of that particular section.

\begin{figure}[H]
\centering
\includegraphics[width=0.8\linewidth]{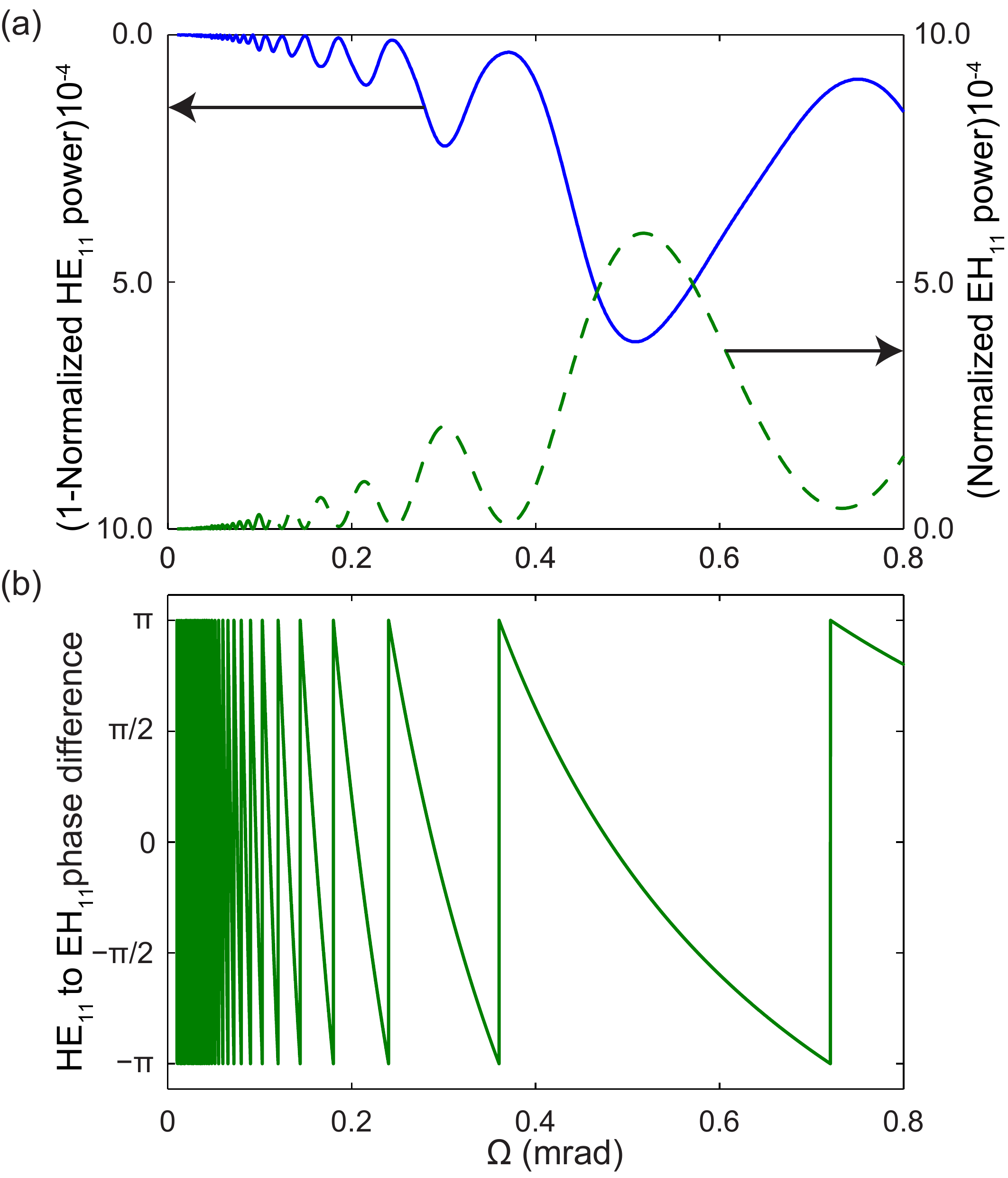}
\caption{ (color online) Transmission of one section (tapering from 25.5 to 23.5~$\mu$m) as a function of angle when the input is the fundamental mode. (a) Amplitude of the fundamental HE$_{11}$ (continuous blue) and the first higher-order mode EH$_{11}$ (dashed green) (b) Phase difference between the fundamental and the first higher order mode.}
\label{fig:onesection}
\end{figure}

Figure.~\ref{fig:onesection} shows the modal evolution in a tapered section when the input is in the fundamental mode. 
When $\Omega$ is small (or the length $L$ is large), the modal evolution is adiabatic and the transmission approaches unity as seen in the plot for the normalized power in the $HE_{11}$ mode in Fig.~\ref{fig:onesection}. When $\Omega$ increases, some energy couples to higher-order modes, and the fundamental mode transmission decreases. For the small angles considered here, Fig.~\ref{fig:onesection}(a) shows energy transfer to one mode only ($EH_{11}$ mode dashed green curve). Energy transfer to other modes ($HE_{12}$ mode and higher) is negligible within the resolution of the plot. The oscillations in the transmission are due to modal dispersion in the fiber, which leads to spatial beating: two modes see different indices of refraction and accumulate a phase difference as they propagate through the fiber (see Sec.~\ref{sec:modes}). The phase accumulation increases and can become large for small angles (or increased fiber length). In the particular situation of Fig.~\ref{fig:onesection}(b) where only two modes beat together, the $EH_{11}$ power reaches local maxima for zero phase differences and local minima for $\pi$ phase differences. The situation can become complex when more than two modes are excited. Consequently, there exist some situations where large intermode energy transfers during the propagation still results in good fundamental mode transmission. Thanks to mode spatial interferences, most of the energy can couple back to the fundamental mode during the propagation. In this case, one relies on interference in the non-adiabatic effects.

\subsection{Genetic algorithm}

We obtain the total transmission $T$ after calculating the projection on the fundamental mode of the full $S$-matrix, given by the product of all $S$-matrices for each section. We want to to find the shortest tapered fiber given a target transmission. For this task, we use the genetic algorithm function from MATLAB to find an optimal solution. This approach is efficient with large problems and allows the use of information of previous runs to improve the computing time in contrast with MonteCarlo methods and other optimizations techniques that use deterministic approaches. Typical parameters for the algorithm are a population size of 500, a crossover probability of 0.7, a mutation probability of 0.025 and a number of generations of 500. The genetic algorithm can probe a large parameter space: for each section, we have calculated 1500 $S$-matrices, for angles that can vary between 10~$\mu$rad and 1.57~rad. We run the algorithm more than 1000 times with different sets of parameters to approach the global minimum.

\subsection{Fully adiabatic fiber}
\label{sec:fullyadiab}

We will define total transmissions greater than $T=0.9990$ as a fully adiabatic fiber. In this section, we investigate fibers with limited intermode energy transfers during the pull. This means that the power contained in the fundamental mode cannot deviate too much from $T$ everywhere in the taper. In this case, the interference between higher order modes plays a minimal role in the final transmission. We benefit from the robustness with respect to variation in parameters that is associated with an adiabatic process. We obtain the most strict condition on the angles that can be used to reach a specific transmission. We run the algorithm with the added condition that the transmission of each small taper section is greater than $T$. That way, we make sure that the fundamental mode power is greater than $T$ at 32 points in the taper. Between those points the fundamental mode power can oscillate, but remains constrained around $T$, ensuring the limitation of intermode energy transfers everywhere in the taper. 

\begin{figure}[H]
\centering
\includegraphics[width=0.8\linewidth]{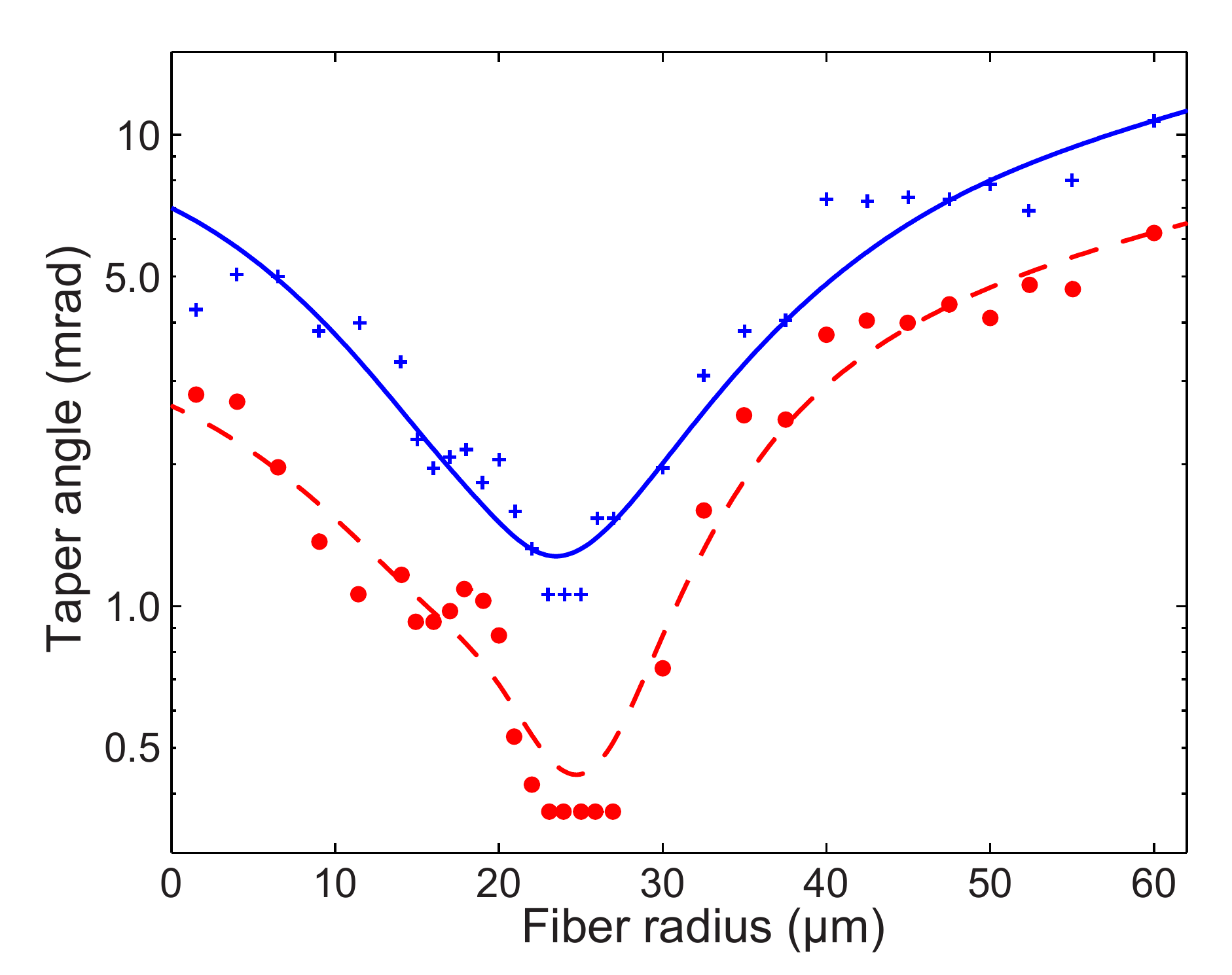}
\caption{(Color online) Optimal adiabatic tapers calculated with the genetic algorithm for $T=99.90\%$ (blue crosses and continuous line) and $T=99.99\%$ (red dots and and dashed line), where the intermode energy transfers are limited. Each marker corresponds to the optimum angle for a section. The lines are guides for the eye.}
\label{fig:adiabthresholds}
\end{figure}

Figure \ref{fig:adiabthresholds} shows results from the genetic algorithm for optimized adiabatic fiber tapers using target transmissions of 99.90\% and 99.99\%. We plot the taper angle as a function of the fiber radius as in Fig.~\ref{fig:adiabaticity}. We observe similar behavior: large taper angles are allowed for large fiber radii, then reach a minimum around the transition region at 20~$\mu$m, before increasing again at smaller radii. For $T=0.9999$, the optimal taper in Fig.~\ref{fig:adiabthresholds} (red dashed curve) shows angles as low as 0.4~mrad, 30 times smaller than the $z_b=z_t$ criteria. The results in Fig.~\ref{fig:adiabthresholds} give precise bounds on adiabaticity, with minimum power transmitted to higher-order modes. This last point insures that this algorithm is insensitive to phase effects: the final transmission is not a consequence of constructive interference between several modes and will be independent of perturbation to the fiber geometry.

\begin{figure}[H]
\centering
\includegraphics[width=0.8\linewidth]{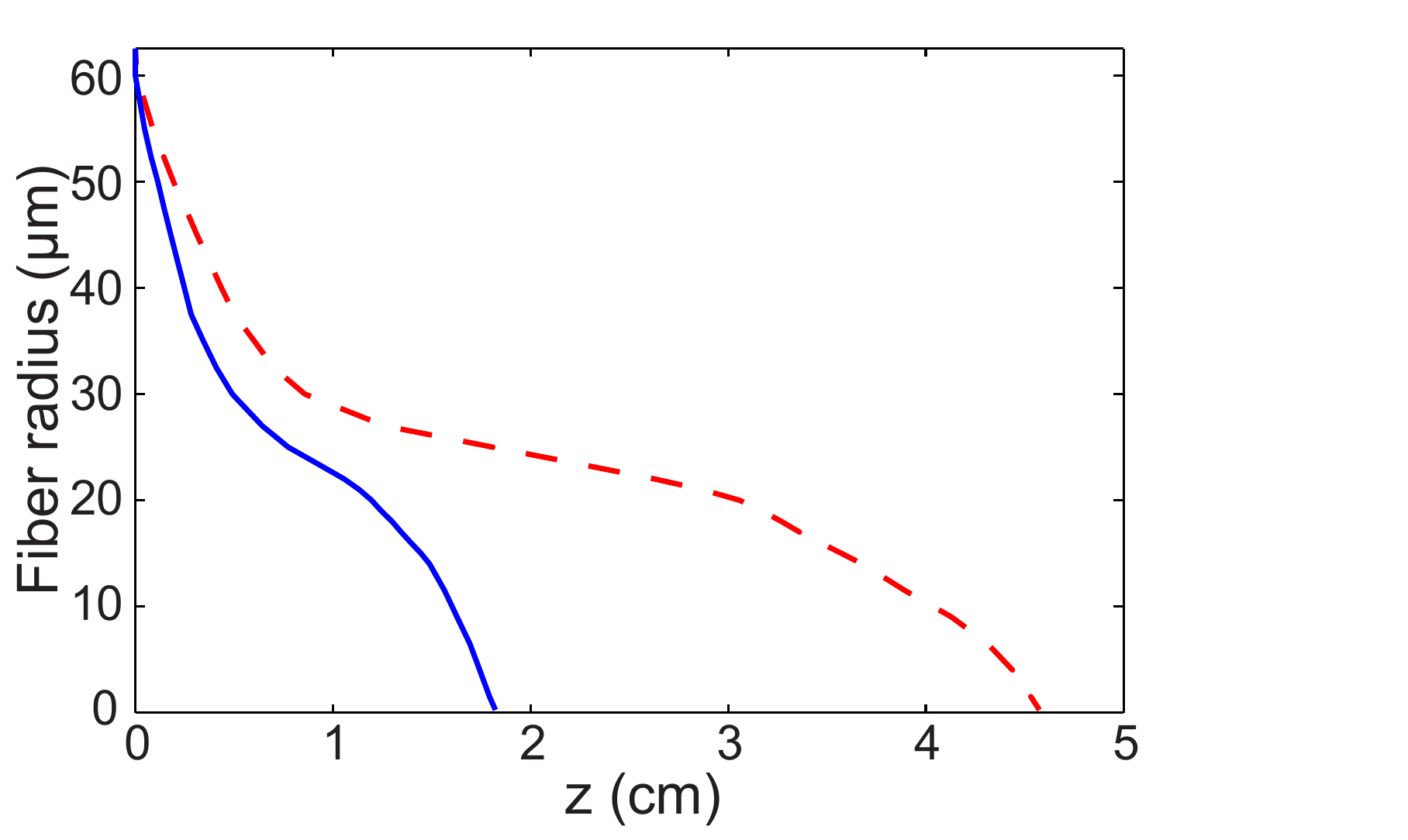}
\caption{(color online) Optimal taper profiles for $T=99.90\%$ (continuous blue line) and $T=99.99\%$ (dashed red line). The profiles are only based on the dots from Fig.~\ref{fig:adiabthresholds} and not on the continuous lines. Note that the horizontal axis scale is in centimeters whereas the vertical axis scale is in microns.}
\label{fig:adiabprofiles}
\end{figure}

Figure~\ref{fig:adiabprofiles} shows the optimized taper profiles corresponding to $T=0.9990$ (blue continuous line) and $T=0.9999$ (red dashed line). Strikingly, for $T=0.9999$ the optimized adiabatic taper is only 4.5~cm long, on the order of typical non-adiabatic tapers lengths produced with a heat-and-pull method \cite{Hoffman2013} (the 2~mrad taper presented Sec.~\ref{sec:transanalysis} is $\approx$~6~cm long and still presents non-adiabaticities). Note however that in Fig.~\ref{fig:adiabprofiles}, $\Omega$ varies continuously as a function of $z$, and can be large at the beginning of the pull. Experimentally, we show below (see Sec.\ref{sec:applications}) that abrupt variations of $\Omega$ during the pull can induce detrimental asymmetries in the taper. With our apparatus, we have precise control of the taper geometry for linear and exponential profiles \cite{Hoffman2013}. Reaching adiabaticity that way would require a linear taper angle $\Omega \approx$~0.5~mrad, and a substantially increased length. One could chose to use smaller clad-fibers \cite{Ravets2013} or to chemically pre-etch fibers, allowing shorter adiabatic tapers with improved handling.

\subsection{Utilizing non-adiabaticity}
\label{sec:shortcuts}

Limiting intermodal energy transfers in a taper to arbitrarly small values is possible, but can be impractical due to large taper lengths. An alternative approach consists of allowing large energy transfers, yet reaching high transmissions by careful design and phase control in the fiber. Section~\ref{sec:modes} shows that different modes interfere together as they propagate in the taper. Taking advantage of this spatial beating, we can design fibers with particular phase combinations that allow high transmission, despite the presence of non-adiabaticities. In this section, we run the genetic algorithm with only a condition on the final transmission ($T\geq0.9999$): intermodal energy transfers in each section is no longer limited. Using this non-adiabaticity, it is possible to produce short high-transmission tapers.

\begin{figure}[H]
\centering
\includegraphics[width=0.8\linewidth]{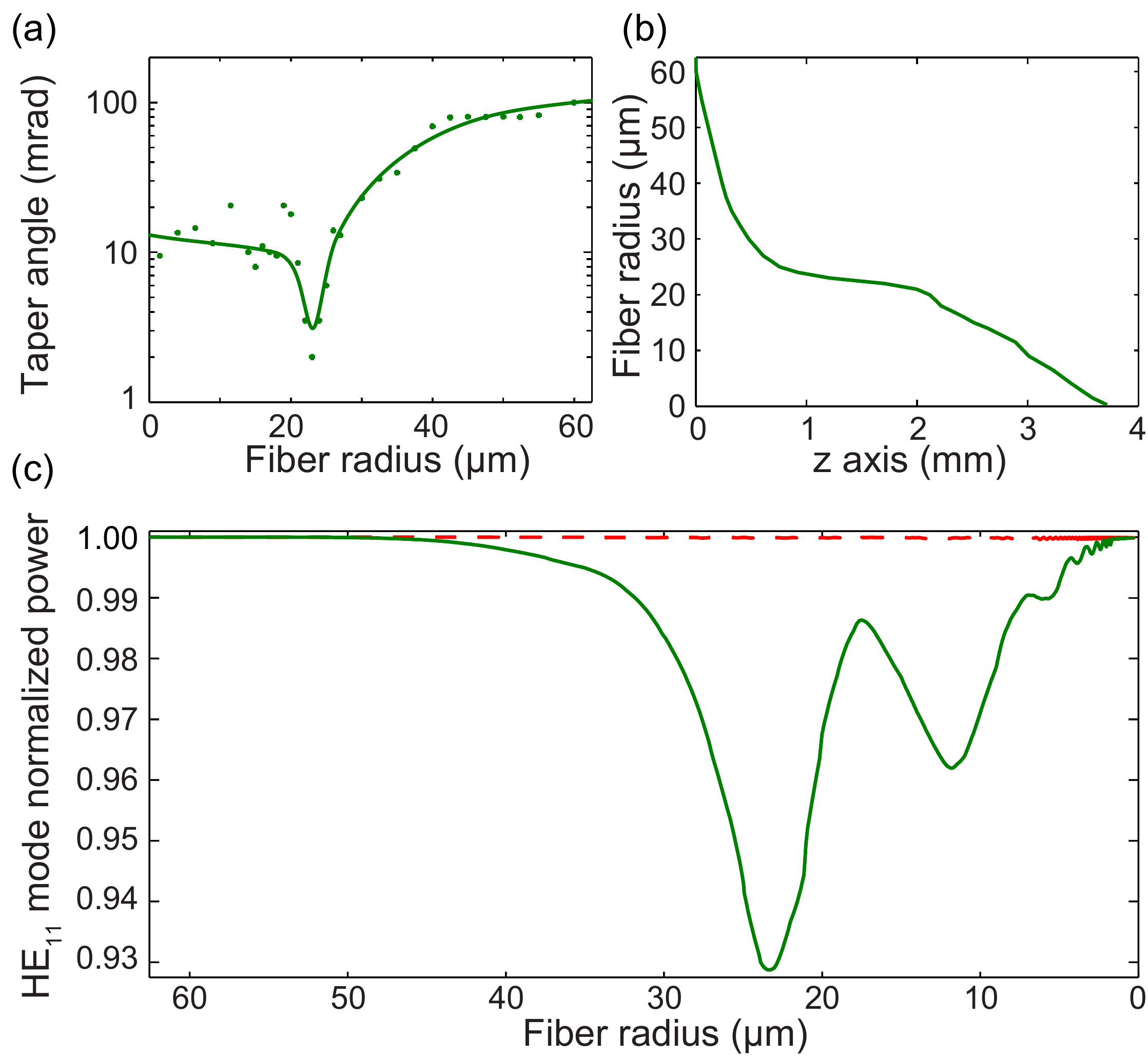}
\caption{(color online) Study of fiber profile for 99.99\% transmission with optimized length given by the genetic algorithm. (a) Taper angles and fiber radius, squares with the continuous line to guide the eye. (b) Fiber radius and fiber length with a final length of 3.7~mm using the results in (a). (c) Fundamental mode transmission as a function of fiber radius for the optimized adiabatic fiber (red dashed line) and the optimized non-adiabatic fiber (green continuous line). 
}
\label{fig:shortcut}
\end{figure}

We calculate the shortest fiber length that has a 99.99\% total transmission in the fundamental mode using the genetic algorithm. Figure~\ref{fig:shortcut}(a) shows that the taper angles allowed here are much larger than the ones presented above in the adiabatic case (Fig.~\ref{fig:adiabthresholds}). At large fiber radii, the taper angle reaches $\approx$~100~mrad. Closer to the transition region, the minimal taper angle can still be as large as 2~mrad. From the the taper angles used here, we know that the fundamental mode is not propagating adiabatically in this taper. Figure~\ref{fig:shortcut}(b) shows the corresponding profile. Figure~\ref{fig:shortcut}(c) shows a 99.99\% transmission (green continuous line) fiber with a 3.7~mm length, a factor of 12 shorter than in the adiabatic case (red dashed line) calculated using FIMMPROP. This greatly reduces the length requirements for high-transmission fibers, which is particularly relevant for our application.

Using FIMMPROP, we model the tapered structures presented in Fig.~\ref{fig:adiabprofiles} and Fig.~\ref{fig:shortcut} by putting together the succession of 32 linear tapered sections obtained with the genetic algorithm. The input is set to 100\% in the $HE_{11}$ fundamental mode and the simulation includes 15 modes of family $l=1$. We calculate the modal evolution (phases and amplitudes) along the taper. In the adiabatic optimized case (red dashed curve Fig.~\ref{fig:shortcut}(b)), we confirm that the power contained in the fundamental mode is close to 99.99\% throughout the taper. Higher-order modes excitations are negligible, and the evolution is adiabatic. Using non-adiabaticity (green curve Fig.~\ref{fig:shortcut}(b)), we observe large energy transfers to higher-order modes. Around $R=23~\mu m$, more than 7\% of the energy has been transferred to higher-order modes. However, using this particular geometry, the resulting phase combinations lead to high-transmission in the fundamental mode.

Non-adiabaticity can lead to high-transmission with shorter tapers, which is particularly useful for taper design. In the rest of the paper, we experimentally study fibers that exhibit this behavior. Exploiting non-adiabaticity needs particular attention because of their sensitivity to mode phases: deviations from the calculated profile might lead to situations where mode interference causes large losses, with less energy ending in the fundamental mode than initially expected. One needs to reproduce the calculated geometry as accurately as possible. As discussed above, producing the taper in Fig.~\ref{fig:shortcut} with a continuously varying angle is not the best option for us, due to the presence of large angles and possible experimental asymmetries. Moreover, this particular taper length (3.7~mm) is too small in comparison to the heating-zone size (0.75~mm in our experiment) to accurately produce such a profile. 
Our typical profiles start with a linear section ($\Omega$ of a few mrad) down to 6~$\mu$m radius, followed by and exponential section down to 250~nm radius. We calculate with FIMMPROP the $HE_{11}$ mode evolution through such a taper ($\Omega =$2~mrad) and show that it benefits from non-adiabatic effects, leading to high-transmission.

\begin{figure}[H]
\centering
\includegraphics[width=0.8\linewidth]{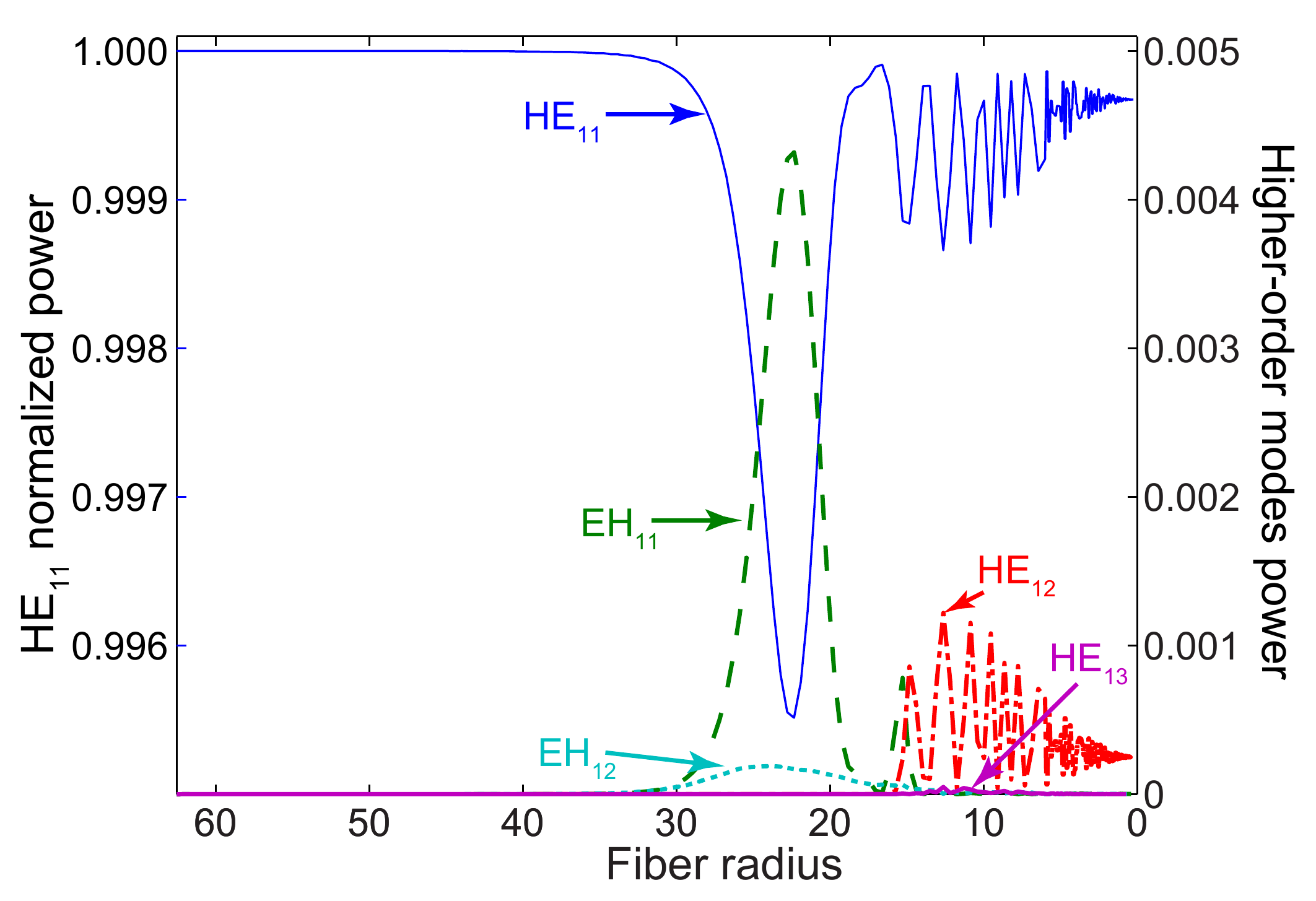}
\caption{(color online) Mode evolution for a 2~mrad linear fiber down to 250~nm radius. During the propagation through the taper, some energy is transferred from the fundamental $HE_{11}$ (blue thin continuous line) to 4 higher-order modes $EH_{11}$ (green long dashed line), $EH_{12}$ (light blue dotted line), $HE_{12}$ (red dashed dot line), and $EH_{13}$ (purple thick continuous line). The final transmission through one taper is 99.97\% on the $HE_{11}$ fundamental mode.}
\label{fig:FIMMPROP2mrad}
\end{figure}

We start by investigating geometries we can produce with good accuracy using our fiber puller. Figure~\ref{fig:FIMMPROP2mrad} shows the transmission of the first few modes of family $l=1$ through a 2~mrad taper. We create a taper with FIMMPROP that reproduces the experimental profile, which has been validated with microscopy measurements \cite{Hoffman2013}. Initially, all the power is contained in the fundamental mode. Around $R=23~\mu$m, $\approx$~0.4\% of the energy is transferred to higher-order modes because of non-adiabaticities (up to $HE_{13}$, the fifth mode of family $l=1$). This illustrates that non-negligible higher-order mode excitations can be observed below the $z_b=z_t$ limit (the taper angle $\Omega=2$~mrad is at least a factor of five below the $z_b=z_t$ limit everywhere in the taper). Those modes beat together, and by the end of the taper, 99.97\% of the energy is transmitted through the fundamental mode. For different taper angles, we observe that our typical tapers benefit from non-adiabaticity (see Sec.~\ref{sec:transanalysis}). If there is still room for optimization, the simplicity of the linear geometry makes it the ideal candidate for our application. 

\section{ANALYSIS OF THE TRANSMISSION SIGNAL}
\label{sec:transanalysis}

\begin{figure}[H]
\centering
\includegraphics[width=0.8\linewidth]{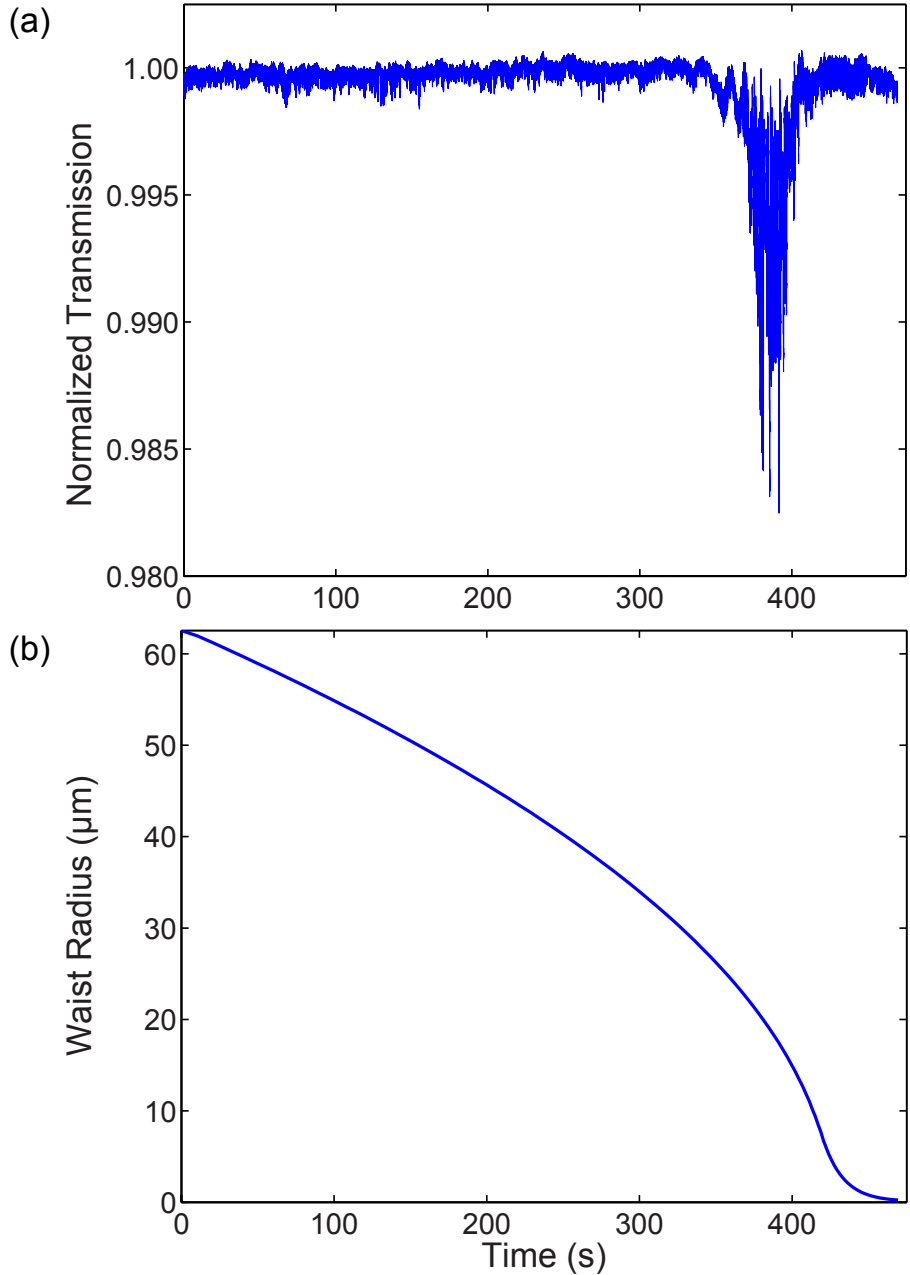}
\caption{(color on line) (a) Transmission through a 2 mrad fiber as a function of time during the manufacturing process. (b) Evolution of the waist radius during the pull, calculated from the algorithm described in \cite{Hoffman2013}. The final radius is 250 nm.}
\label{fig:trans2mrad}
\end{figure}

We evaluate the quality of a pull by monitoring the transmission of a few mW from a 780.24 nm laser through the fiber during the process. We normalize the signal to remove fluctuations of the laser intensity. Figure~\ref{fig:trans2mrad}(a) shows a typical transmission as a function of time for a successful 2 mrad pull. The transmission and normalization fiber outputs are connected to two Thorlabs DET10A photodetectors, which deliver a signal to a SR570 low-noise differential preamplifier from Stanford Research Systems. A Tektronix DPO7054 digital oscilloscope set on high resolution mode and sample rate of 10-20 ksample/s records the data. The fiber is thinned during the pull, and as its radius decreases, we observe different notable features in the transmission signal. Figure~\ref{fig:trans2mrad}(b) shows the relation between time and radius for the particular pull of Fig.~\ref{fig:trans2mrad}(a) calculated using the algorithm for fiber pulling that was validated in \cite{Hoffman2013}, with a deviation from the experimental measurements lower than 8\% at all diameters .

\subsection{Single mode section}
\label{sec:SingleMode1}

The fiber is initially close to being single mode ($V \approx 2.45$) at the light wavelength we use to measure the transmission. We carefully launch the fundamental mode with a 1.27 cm radius of curvature mandrel wrap, to filter higher-order modes from the initial launch. During the first 100 seconds (down to 25~$\mu$m radius), we observe a constant transmission. A 2-mrad taper is completely adiabatic in this region (see Fig.~\ref{fig:adiabaticity}). The fundamental mode is confined to the core and does not interact with any other mode.

\subsection{From single mode to multimode}
\label{sec:Multimode}

\begin{figure}[H]
\centering
\includegraphics[width=0.8\linewidth]{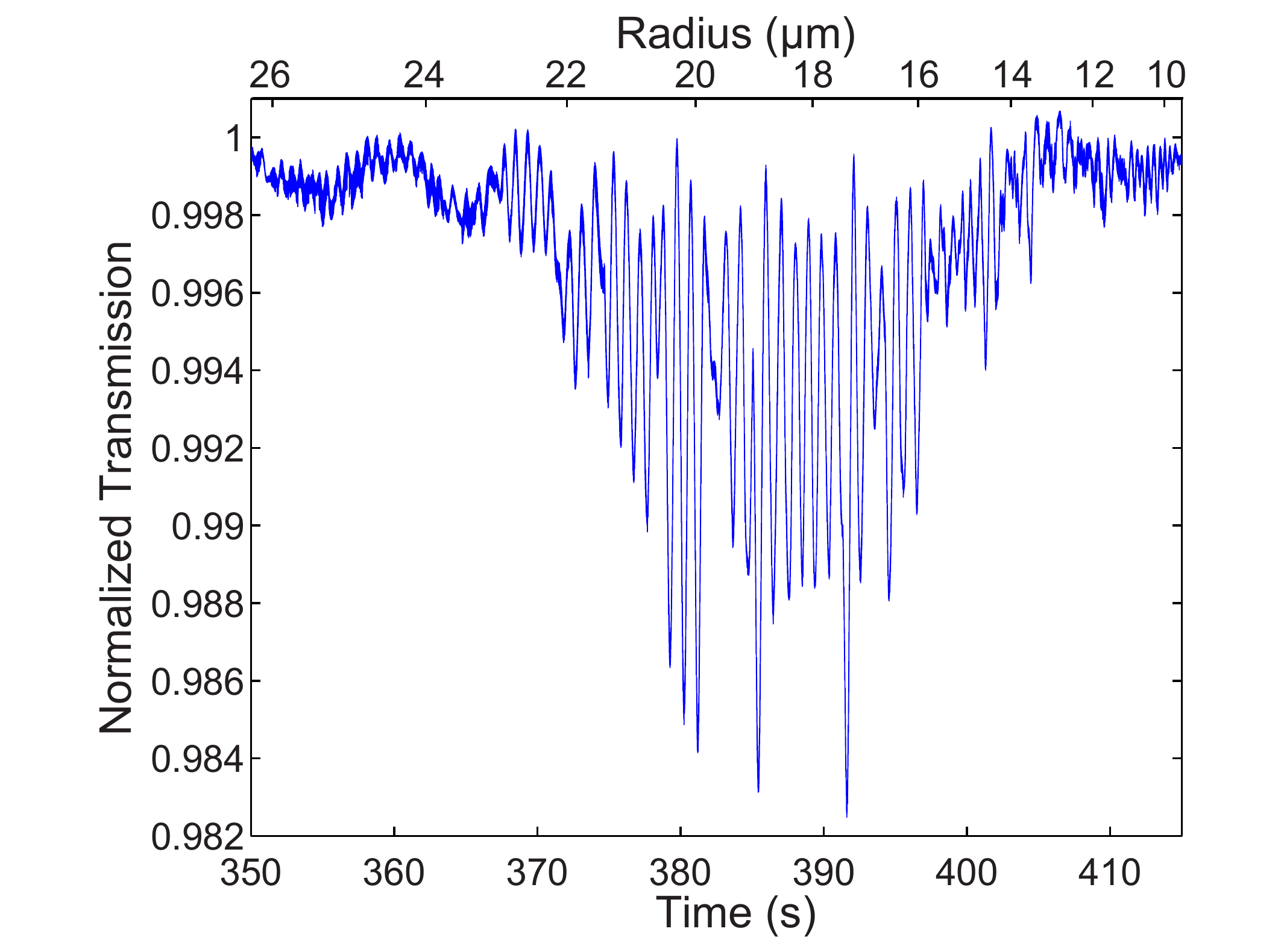}
\caption{Zoom of the transmission of the 2 mrad pull showed in Fig.~\ref{fig:trans2mrad}, when the radius of the waist is near 20~$\mu$m. We see oscillations in the transmission signal, due to the beating between the fundamental mode and higher-order modes excited at a radius of 20 microns. The top vertical scale is the fiber radius at the waist.}
\label{fig:zoom}
\end{figure}

As the fiber radius decreases, the fundamental mode effective index approaches the cladding index of refraction (see Fig.~\ref{fig:dispersion}). The fiber core becomes too small to support the fundamental mode, which progressively leaks into the cladding to become guided by the cladding-to-air interface. The point where the fundamental mode leaks into the cladding is $n_{eff}^{HE11} = n_{clad}$, at $R=19.43$~$\mu$m. At that point, the waveguide is so large in comparison to the wavelength of the light that the fiber is multimode ($V \approx 170$). The dispersion relation curves of all the modes are close to each other (see Fig.~\ref{fig:dispersion}(a)), and the modes can easily interact. Figure~\ref{fig:adiabthresholds} shows that the tapering angle has to be smaller than 0.3~mrad in order to be adiabatic in that region. The transmission signal therefore shows mode beating (see Fig.~\ref{fig:zoom}).

Energy transfers to higher-order modes occur during the transition from core to cladding because of this non-adiabaticity. For a cylindrically symmetric fiber, such a transfer of energy is only possible from the fundamental mode to other modes of order $l=1$ (by symmetry). Once they have been excited in the fiber, those modes coexist and propagate together with different propagation constants, given by the dispersion relation curves (green curves in Fig.~\ref{fig:dispersion}). The optical path length inside the fiber is:
\begin{equation}
\label{eq:OPL}
(L)_n = \int_{fiber} n_{eff}^n(z) dz.
\end{equation}
Equation~(\ref{eq:OPL}) shows that different modes accumulate a phase difference. The modulation observed in the transmission signal around radius 20~$\mu$m is a signature of the presence of higher-order modes beating together.

\begin{figure}[H]
\centering
\includegraphics[width=0.8\linewidth]{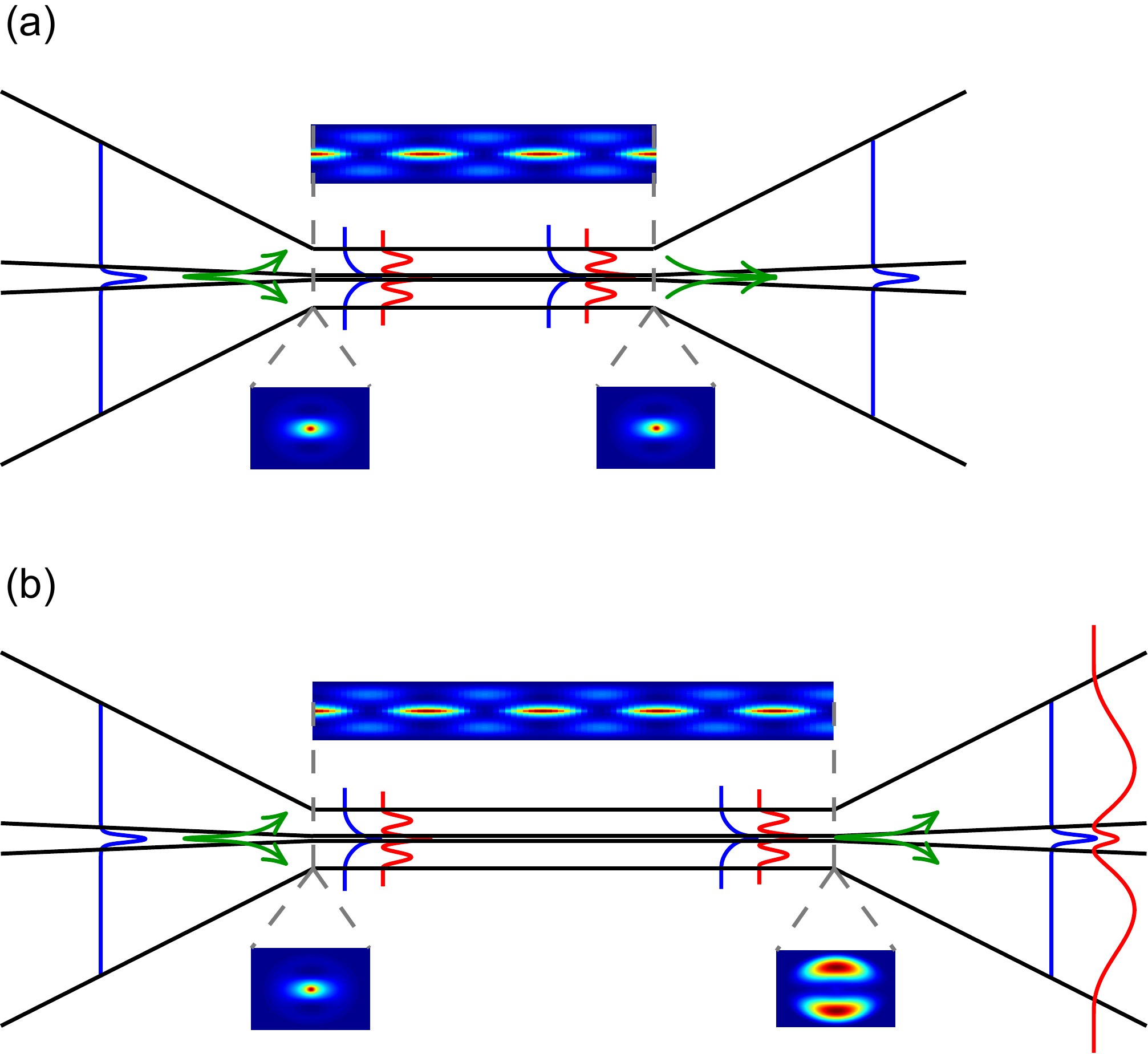}
\caption{(color online) Schematic of the modal evolution in the transition region. All the power is initially contained in the fundamental mode (blue profile). When the core of the fiber becomes too small compared to the wavelength, the light escapes into the cladding (green arrows) and some higher-order modes can be excited (red profile). The radius of the waist is equal to 20~$\mu$m, so that the excited modes do not experience any cutoff as they propagate through the waveguide. (a) The length of the fiber is an integer number of beating lengths. (b) Length of the fiber not an integer of beating lengths. The mode profiles were calculated with FIMMPROP.}
\label{fig:beating}
\end{figure}

The beating signal is related to the relative phases when light couples back into the fiber core as the $R$ increases in the second taper. When $R$ reaches 19.43~$\mu$m in the second taper, energy couples back into the core. Although the two tapers are identical, the presence of beating between modes breaks the symmetry (see Fig.~\ref{fig:beating}). Depending on the fiber length, the phase accumulation between the modes leads to a different field distribution entering the core at 19.43~$\mu$m. The fraction of energy that can couple back into the core depends on the field distribution at this point. If the modes travel through an integer number of beat lengths, the field distribution returns to its initial input. The reciprocity theorem implies that all the energy couples back into the core. If the modes experience a non-integer number of beat lengths, the field distribution is different from what it was initially and only a fraction of energy can couple back into the core: the rest of the energy couples to cladding modes. The cladding light is not detected since we filter the higher-order modes placing a mandrel wrap in front of the detector. At the fiber output, we therefore only observe on the detector the light that coupled back into the fiber core.

\subsection{Single mode again}
\label{sec:SingleMode2}

As we continue to thin the fiber, the modes' effective indices approach the air index of refraction. When $R$ (equivalently, the $V$-parameter) becomes small enough, the excited modes cut off and couple to radiation modes in air. Meanwhile, the fundamental mode's effective index asymptotically approaches $n_{air}$ without reaching a cutoff. A small enough fiber can consequently be single mode again after all the higher-order modes cutoff. For the SM800 fiber, the single-mode cutoff occurs at 300~nm radius. After this cutoff, we do not see any beating anymore, and the transmission is steady again: we measure a transmission of 99.950(23)\% where the dispersion of the distribution is $5.8 \times 10 ^{-3}$ and the dispersion on the mean is $1.2 \times 10^{-5}$. Possible systematic effects related to the fiber cleanliness and the detectors and amplifiers long term stability prevent us from giving a better bound than 0.023\% to the measured uncertainty in the transmission, but $T$ is close to unity, both in the measurement and in the simulation. Note that the simulation Sec.~\ref{sec:shortcuts} looks at the propagation through a single taper. In the present case, light goes through two tapers, explaining why the measured transmission is slightly smaller than the simulated one.

\section{SPECTROGRAMS}
\label{sec:spectrogram}

\begin{figure}[H]
\centering
\includegraphics[width=0.8\linewidth]{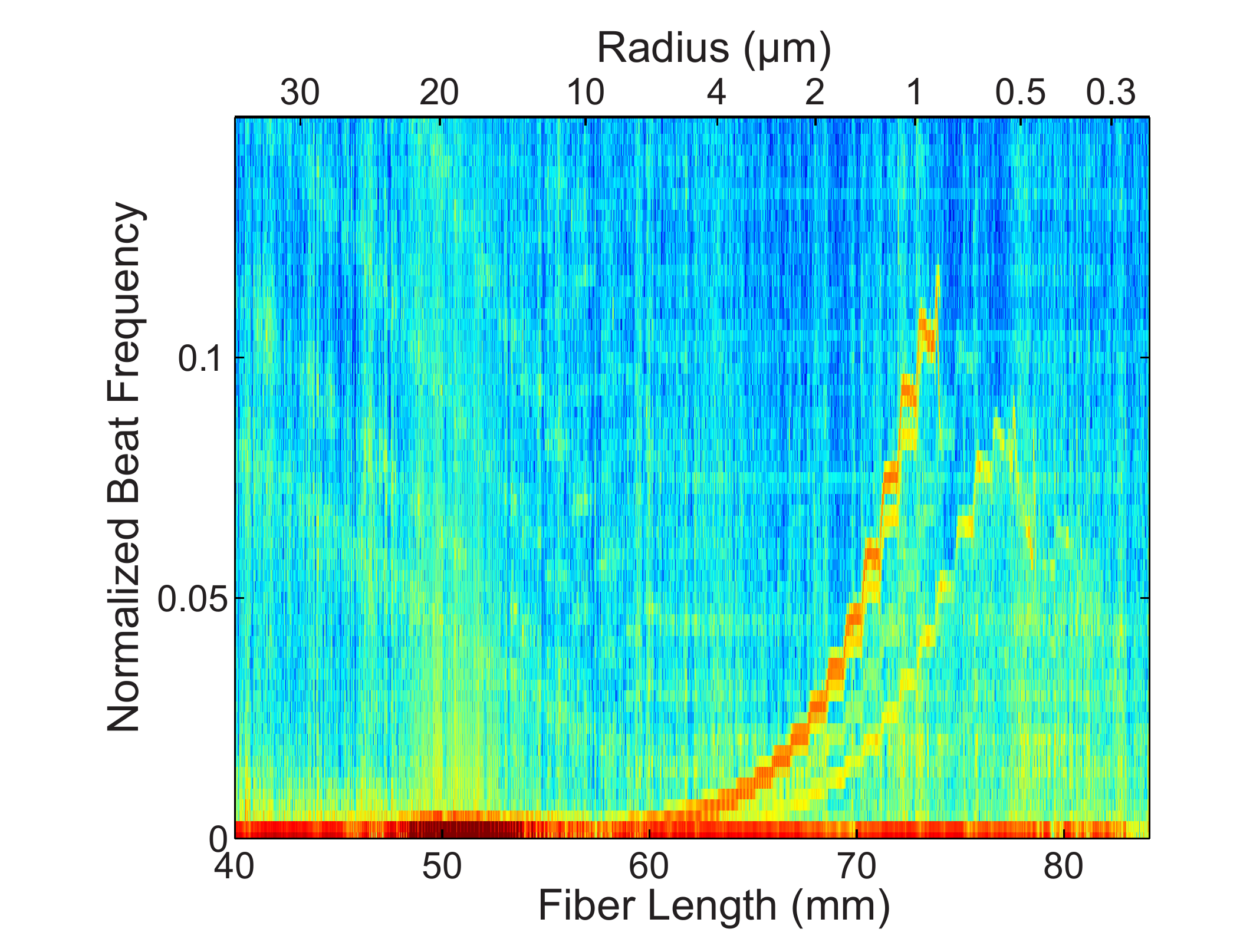}
\caption{(color online) Transmission spectrogram of a 2 mrad pull (see the time evolution in Fig. \ref{fig:trans2mrad}) as a function of the stretch $L$ of the fiber, showing the chirp of the beating frequency and the abrupt end of the beating. The top vertical scale shows the waist radius calculated from the algorithm. The colormap corresponds to the power spectral density (PSD).}
\label{fig:spectrogram2mrad}
\end{figure}

We extract the evolution of the frequencies contributing to the beating process as a function of pull-time using spectrograms, which plot local, windowed Fourier transforms of the transmission signal as a function of time. We use the spectrogram function in MATLAB with a window of 8192 points and an overlap of 7000 points (See Fig.~ \ref{fig:spectrogram2mrad} for an example of a spectrogram of the transmission from Fig.~\ref{fig:trans2mrad}(a)). The modulation in the transmission does not have a single frequency. The frequency is chirped for various reasons. First, the stretch of the fiber is not a linear function of time. Its form depends on the chosen pulling parameters, and can be calculated using our algorithm. Second, the propagation constants of the modes are not only radius-dependent but the way they evolve also depends on the mode. The difference between two curves varies as a function of $R$, which means that the phase does not accumulate at a constant rate.

Each curve in the spectrogram signals the interaction between two modes beating at a given frequency. They all appear when the fiber enters the multimode regime ($R \approx 19.43$~$\mu$m). The presence of these curves indicates non-adiabaticities in the pull. The curves terminate before the end of the pull, at a point that corresponds to the cutoff of one of the two beating modes. We now have the task to identify which modes are excited, how they are excited, and if there is a way to suppress their excitation. Given the specificity of the phase accumulation for a couple of modes, it is possible to label the modes excited during the pull and use the spectrogram as a diagnostic to evaluate the adiabaticity and symmetry of the fibers.

\subsection{Modeling the pull}
\label{sec:model}

The phase accumulation between two modes is a function of their optical path length, which depends on the geometry of the fiber at a time $t$ (see Eq.~(\ref{eq:OPL}) above). When the stretch at that time is equal to $L$, the phase accumulation between two modes is:

\begin{equation}
\Phi_{i,j} (L) = \int_0^L \left[ \beta _ {i}(r(z)) - \beta _ {j}(r(z)) \right] \, \mathrm{d} z,
\end{equation}

with spatial frequency $K$ \cite{Orucevic2007}

\begin{equation}
K_{i,j} (L) = \frac{1}{2 \pi}\frac{d \Phi _{i,j}}{dL}.
\end{equation}

\subsection{Identifying the modes}
\label{sec:identification}

\begin{figure}[H]
\centering
\includegraphics[width=0.7\linewidth]{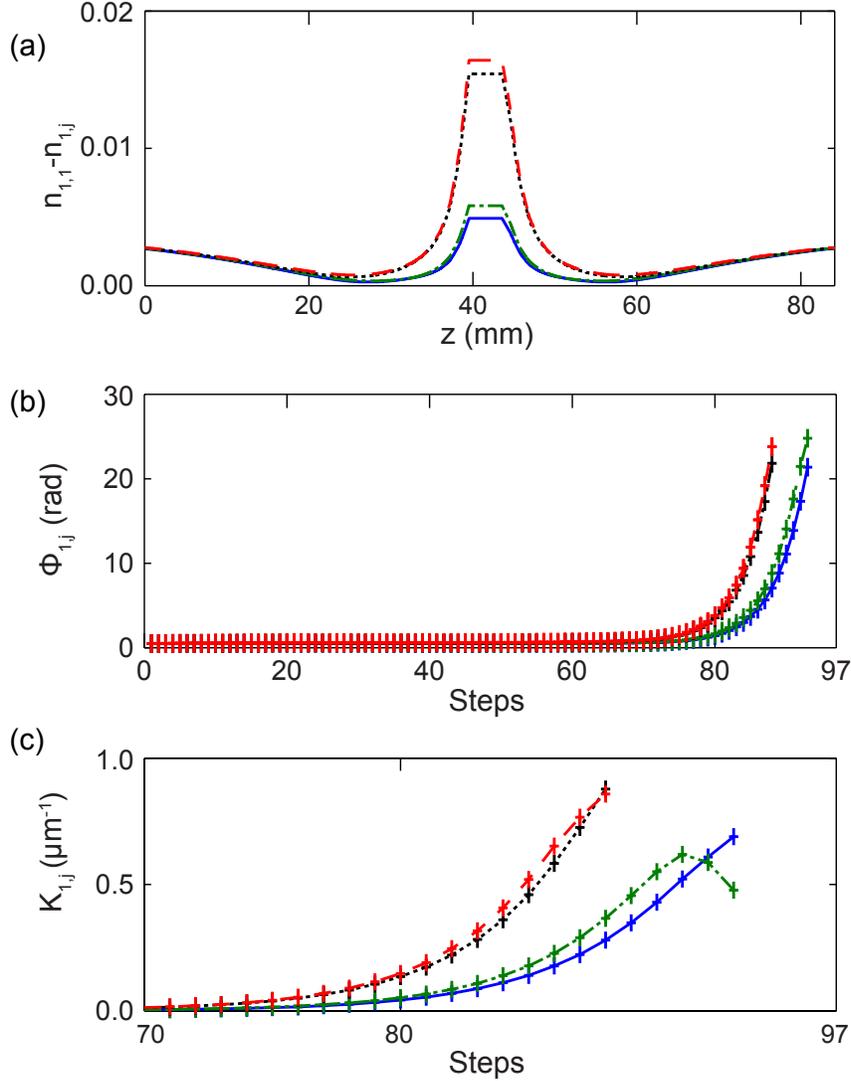}
\caption{(color online) Study of the differences between the fundamental mode and the first four excited modes of family one. (a) $\Delta \beta _ {1,j}$ as a function of length along the fiber axis (difference between the indices of refraction at step 75). (b) Phase accumulation $\Phi_{1,j}$ as a function of step. (c) Spatial frequency $K_{1,j}$ of the beating as a function of step. The lines (long dashed red, continuous blue, short dash black and long-short dash green) join the calculated points.}
\label{fig:Kijplot}
\end{figure}

We know the profile at the end of a step from the pulling algorithm described in \cite{Hoffman2013}. We use the dispersion relations obtained with FIMMPROP to calculate the differences, $\Delta \beta _ {i,j}$, in propagation constants for mode $i$ and mode $j$. By integrating numerically $\Delta \beta _ {i,j} (z)$ at each step, we obtain $\Phi_{i,j} (L)$. A numerical differentiation of $\Phi_{i,j}$ with respect to $L$ gives us the evolution of the spatial frequency as a function of step (see Fig.~\ref{fig:Kijplot}). From our simulation of the pull, we know the stretch as a function of time, and we can plot the evolution of the spatial frequency as a function of time.

We calculate the spatial frequency for a thousand pairs of modes with different radial symmetry ($l=$1 to 6) and azimuthal order ($m=$ 1 to 20), and we map them on the spectrogram. We can then identify and label the curves observed on the spectrogram by looking for their overlap with the experimental curves. We get an excellent matching without any scale adjustments.

\begin{figure}[H]
\centering
\includegraphics[width=0.8\linewidth]{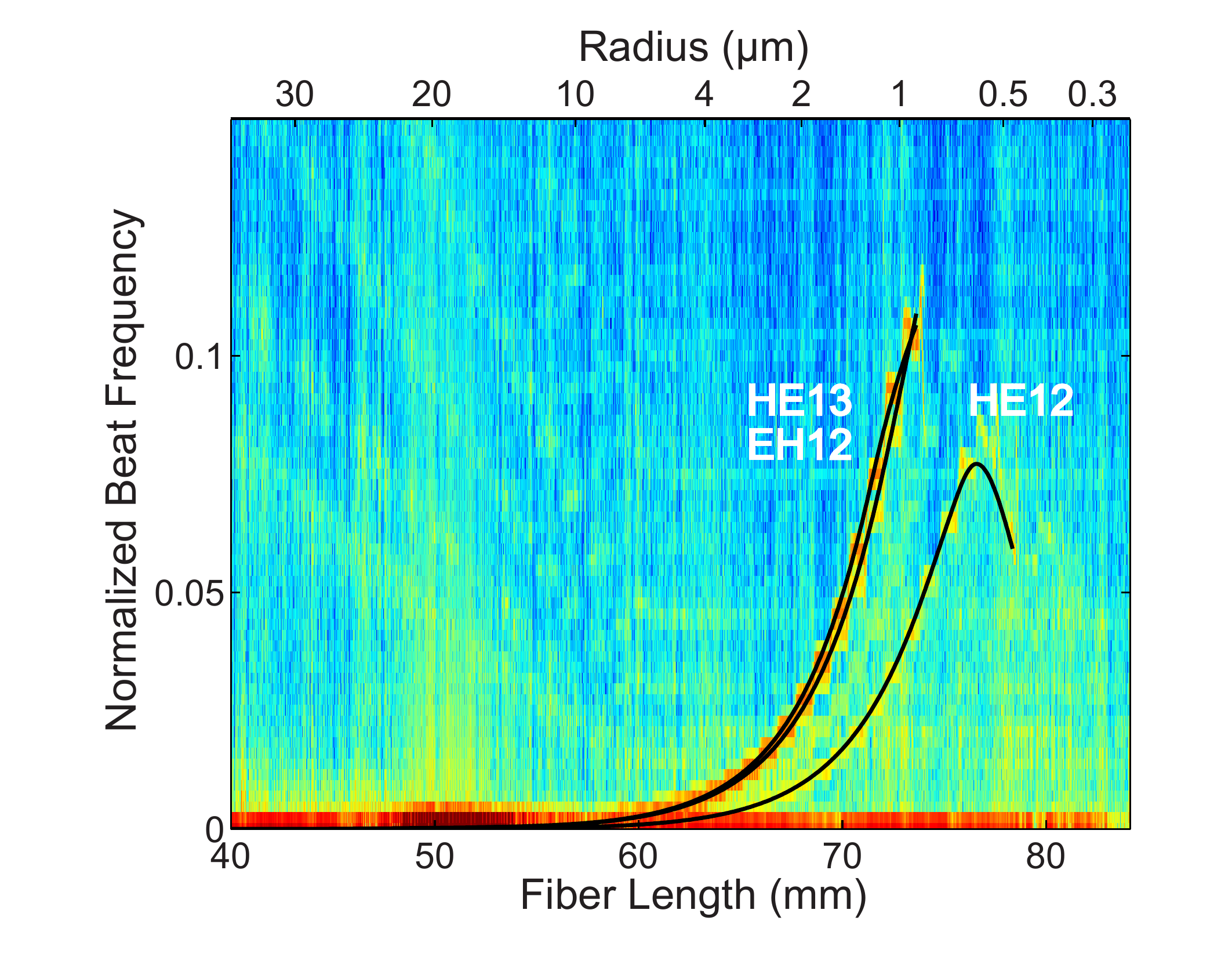}
\caption{(color online) Identification of the modes beating for the 2 mrad tapered fiber spectrogram of Fig.~{\ref{fig:spectrogram2mrad}}. Only modes from family one are beating.}
\end{figure}

We can identify in the spectrogram (2-mrad tapered fiber) the signature of beating between four modes. All those modes have the symmetry $l=1$ (see Fig.~\ref{fig:dispersion}) as expected for a cylindrically symmetric fiber. We observe higher-order mode excitation up to the fifth mode of family $l=1$, which is consistent with the simulations (sec.\ref{sec:shortcuts}). The total $HE_{11}$ transmission is 0.9995, meaning that 0.05\% of the energy has been transferred to the other modes. We suppose here that all the losses come from the transfer of power to other modes. This power is radiated in air when those modes reach cut off. The contribution of other losses like Rayleigh scattering are expelted to be much smaller. The power spectral density (PSD), which defines the colormap in a spectrogram, gives a representation of how the remaining power is distributed between the higher-order modes as a function of time. By plotting the PSD at different times, we evaluate the power contained in each branch contributing to the beating. Below $R=4~\mu$m, those contributions are almost constant, and the higher-order mode relative power is distributed as follow: $5.5 \pm 0.5\%$ in $HE_{12}$, $9 \pm 0.5\%$ in $EH_{12}$ and $85.5 \pm 0.5\%$ in $HE_{13}$. Note that we only resolve the beating between the fundamental mode and one excited mode. The beating between excited modes exists, but this second order effect is too weak to be visible in the spectrogram.

\section{APPLICATION: THE QUALITY OF THE PULL}
\label{sec:applications}

We can use the spectrogram analysis to design and diagnose its quality while pulling a fiber. The number of modes excited and which modes are excited give us information about the adiabaticity, asymmetries and the quality of the fiber after the pull.

\subsection{Multi-angle taper}
\label{sec:algoparam}

The beating amplitude and higher-order modes excitations seen in Fig.~\ref{fig:trans2mrad} and Fig.~\ref{fig:spectrogram2mrad} show that the angle of tapering near the critical region at 19.43~$\mu$m, is non-adiabatic. A shallower taper angle around that region could lead to a more adiabatic transition. Following this idea, we study a fiber with a 2 mrad angle until a radius of 20~$\mu$m, and then decrease the angle to 0.75~mrad. After $R = 6$~$\mu$m, the pull is exponential down to $R = 250$~nm. 

\begin{figure}[H]
\centering
\includegraphics[width=0.8\linewidth]{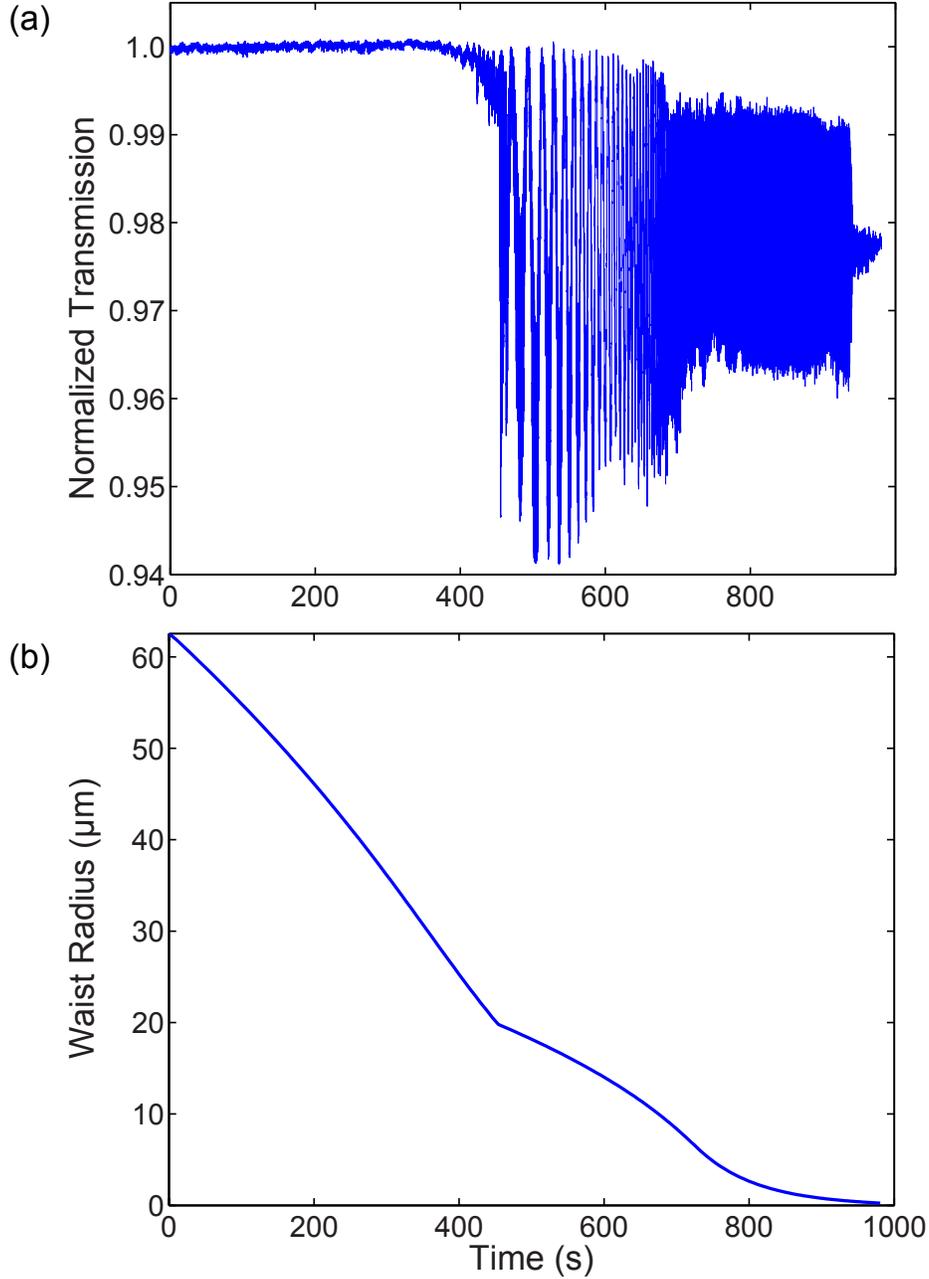}
\caption{(a) Normalized transmission through the fiber as a function of time during the manufacturing process. (b) Evolution of the radius of the waist during the pull. Based on the algorithm, we initially taper the fiber with a 2 mrad angle until a radius of 20~$\mu$m; the angle changes to 0.75~mrad until the radius of the fiber is equal to 6~$\mu$m, where the radius exponentially decreases down to 250~nm.}
\label{fig:transmultiangle}
\end{figure}

We see that the transmission at the end of the pull is only 97.850\% from Fig.~\ref{fig:transmultiangle}. 
This corresponds to a transfer of energy to the higher-order modes larger than 3\%, a factor of sixty worse than in the linear 2~mrad pull (Fig.~\ref{fig:trans2mrad}). The beating amplitude is much larger than in the 2~mrad case. This is surprising since this pull is designed to be more adiabatic, and simulations with FIMMPROP confirm that we still expect a transmission $T~\geq$~99.90\%.

\begin{figure}[H]
\centering
\includegraphics[width=0.8\linewidth]{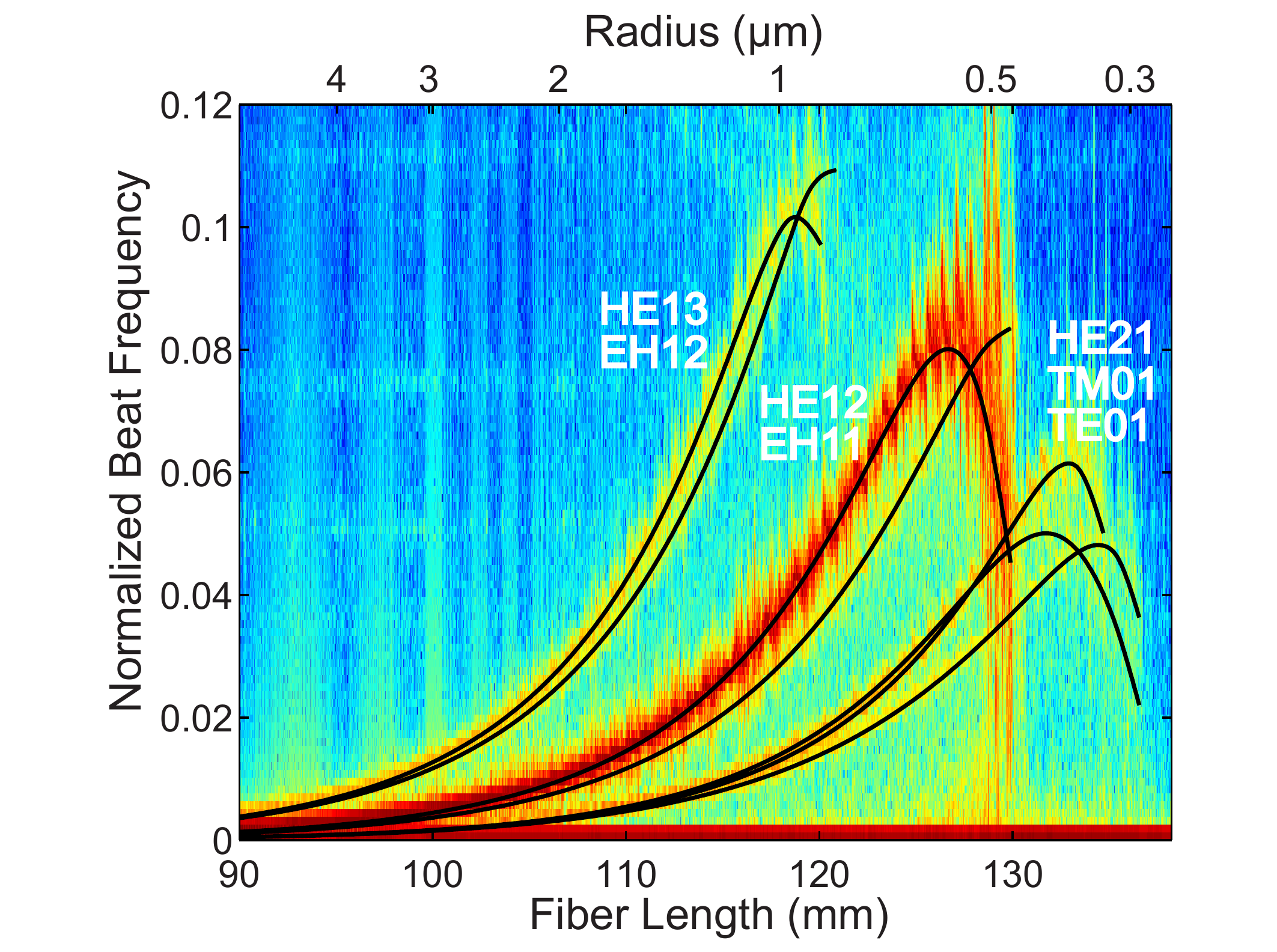}
\caption{Spectrogram of the transmission data shown in Fig.~\ref{fig:transmultiangle}. The solid black curves are the one given by the simulation. The modes are labeled on the figure. The total transmission in the fundamental is 0.97850. For $R \leq 2~\mu$m we calculate from the PSD that the remaining energy is distributed between seven higher-order modes as follow: $TE_{01}$ (0.08\%), $TM_{01}$ (0.05\%), $EH_{11}$ (0.35\%), $EH_{12}$ (0.05\%), $HE_{12}$ (98.4\%), $HE_{13}$ (0.2\%) and $HE_{21}$ (0.87\%).}
\label{fig:spectroverlap2to075}
\end{figure}

\subsection{Tracking asymmetries}

\begin{figure}[H]
\centering
\includegraphics[width=0.8\linewidth]{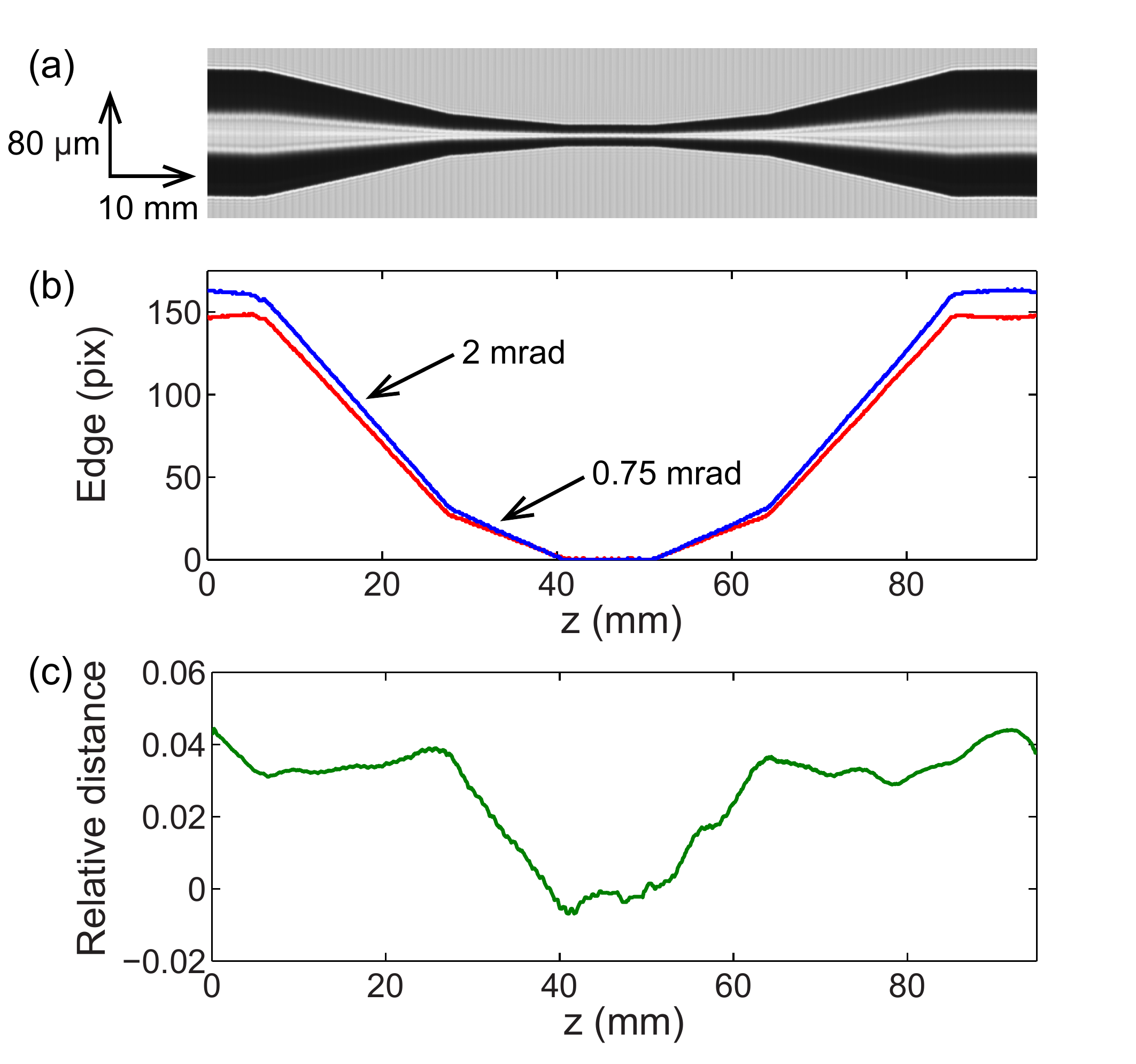}
\caption{(color online) Study of the asymmetry of a pulled fiber. The fiber has a 10~$\mu$m radius with an angle change from 2 mrad to 0.75 mrad at 20~$\mu$m. (a) 100 images taken with an optical microscope stacked and horizontally compressed to enhance any asymmetries. (b) Profile of the bottom edge (blue curve) and top edge (red curve) of the fiber. The abrupt change in angle at 20~$\mu$m introduces an asymmetry at this radius. (c) Relative difference between the two edges (normalized by the diameter of the fiber) as a function of $z$. \label{fig:asymmetry} }
\end{figure}

The spectrogram analysis in Fig.~\ref{fig:spectroverlap2to075} shows excitation to the $TE_{01}, TM_{01},$ and $HE_{21}$ modes, which do not belong to the family of the fundamental mode. The largest transfer is still to the same family, with a different distribution. Coupling to other families should not be observed for a fiber with cylindrical symmetry. This suggests that our multiple angle tapers introduce some asymmetries in the fiber. We imaged the fiber using an optical microscope near the angle change regions (see Fig.~\ref{fig:asymmetry}) to further investigate the decrease of transmission.

Figure~\ref{fig:asymmetry} shows that the bottom angle of the fiber exceeds the top angle. Although the measured diameters are as expected, superimposed plots of the top and bottom edges show that there are imperfections around the transition. We observe a peak at the transition radius ($R \approx$~20~$\mu$m) in the distance between the edges. We believe that the excitation of higher-order modes at this radius is a consequence of this asymmetry. We do not observe the same imperfection around the transition region for a 2~mrad flat fiber. The abrupt change in angle exacerbates imperfections in the pulling process by introducing some asymmetries. This results further support that single-angle linear tapers are good candidates for our application. Further increasing adiabaticity would require to decrease $\Omega$, leading to large taper lengths. Because of geometrical and handling constraints, we find it ideal to work with 2~mrad tapers. To work with steep and multiple angles might require a smaller flame or a more symmetric heating.

\section{UNDERSTANDING THE LOSSES}
\label{sec:thelosses}
\subsection{losses}
Understanding the losses in nanofibers is important for our future applications \cite{Hoffman2011,Hoffman2013}, which require knowledge of such photon loss. We identify two main loss mechanisms that contribute to the final losses: coupling to higher-order modes through non-adiabaticities and scattering around the waist of the nanofiber \cite{Snyder1983}. Systematic effects like the presence of impurities on the fiber surface, or asymmetries in the pull, enhance the losses through those mechanisms.

\subsection{Coupling to higher-order modes}

We have observed and characterized in this paper the effect of non-adiabaticities in the taper. Their presence induces energy transfers to higher-order modes. As we reach the single mode regime, those higher-order modes cut off. They can not be guided by the fiber anymore, and they diffract out as radiative modes. In a plane transverse to the fiber, one can observe a characteristic diffraction pattern further supporting that this effect is most the important for the pulls considered in this study. 

\subsection{Rayleigh scattering}

Rayleigh scattering is present 
in any glass, leading to scattering of light and attenuation in the transmitted signal \cite{Snyder1983}. The attenuation coefficient for fused silica is small at a wavelength of 780.24~nm. By imaging the fiber, it is possible to directly observe the scattering. Experimentally, it is particularly visible on the fiber waist, but remains of the order of 3~dB/km, justifying the fact that we neglected it in this paper. 

\subsection{Systematic effects}

The transmission varies drastically with the surface state of the fiber. When the fiber is initially dirty, the spectrogram analysis shows the excitation of more modes corresponding to more losses. We attribute this to the presence of impurities on the surface of the fiber at the beginning or during the pull. A dust particle on the fiber waist leads to losses through coupling to higher-order modes or scattering. The cleanliness of the fiber is critical before and during the pull. Such imperfections are avoidable by properly cleaning the fiber and imaging the fiber prior to a pull as explained in \cite{Hoffman2013}. All the pulls presented in this paper were done after applying the cleaning procedure described in \cite{Hoffman2013}.

\section{CONCLUSION}
\label{sec:conclusion}
We have demonstrated our ability to produce ultra-low loss optical nanofibers. Reaching high transmissions is important for many nanofiber applications. We have described an algorithm that calculates the optimum taper length for a given transmission, or equivalently the optimum transmission for a given taper length. This new approach concerning adiabaticity in tapered fibers gives more precise bounds than the traditional adiabaticity condition, which helps for the design of a suitable taper geometry. We show that in our experiments, the transition from the single-mode regime to the multimode regime is non-adiabatic, inducing excitations of higher-order modes during the tapering. Having a good control of the taper geometry is crucial for limiting losses due to those excitations. 

The propagation of different modes during the pull leads to a characteristic beating pattern in the transmission. Plotting the spectrogram of the transmission signal and using a model of the fiber pulling, we are able to identify the modes excited during the pull. This gives information for the analysis of the quality of a fiber and the understanding of loss factors, that will help in the manufacturing of even more adiabatic fibers.

\section*{Acknowledgments}
We would like to thank Prof. A. Rauschenbeutel for his interest and support on this project. This work was funded by the National Science Foundation through the Physics Frontier Center at the Joint Quantum Institute, and the Army Research Office Atomtronics MURI. S. R. acknowledges support the Fulbright Foundation.






\end{document}